\documentclass[12pt]{article}
\usepackage[top=2.54cm,left=2.54cm,right=2.54cm,bottom=2.54cm]{geometry}
\usepackage{graphicx} % Required for inserting images
\usepackage{natbib}
\usepackage{amsmath}
\usepackage{latexsym, amsbsy, amssymb, amsthm}
\usepackage{booktabs}
\usepackage{bm}
\usepackage{bbm}
\usepackage{tikz}
\usepackage{multirow,verbatim}
\usepackage{color}
\usepackage{xcolor}

\DeclareMathOperator*{\argmin}{arg\,min}
\def\var{\mathrm{var}}

\RequirePackage[colorlinks,citecolor=blue,urlcolor=blue,linkcolor=blue]{hyperref}
\usepackage{algorithm}
\usepackage{algorithmic}

\newcommand{\indep}{\;\, \rule[0em]{.03em}{.6em} \hspace{-.25em}
\rule[0em]{.65em}{.03em} \hspace{-.25em}
\rule[0em]{.03em}{.6em}\;\,}

\def\eop{\hfill $\Box$}

\usepackage{graphicx}
\usepackage{subcaption}

\usepackage{xr}
\def\spacingset#1{\renewcommand{\baselinestretch}%
{#1}\small\normalsize} \spacingset{1}

\date{}

\theoremstyle{plain}

\newtheorem{theorem}{Theorem}[section]
\newtheorem{lemma}[theorem]{Lemma}
\newtheorem{proposition}{Proposition}[section]

\theoremstyle{definition}

\newtheorem{assumption}{{Assumption}}

\begin{document}

\def\real{\mathbb R}
\newcommand\independent{\protect\mathpalette{\protect\independenT}{\perp}}
\def\independenT#1#2{\mathrel{\rlap{$#1#2$}\mkern2mu{#1#2}}}

\def\boe{\begin{enumerate}}
\def\eoe{\end{enumerate}}
\def\dom{\mathrm{dom}}
\newtheorem{corollary}{{\bf Corollary}}
\newtheorem{conjecture}{{\bf Conjecture}}
\newtheorem{problem}{{\bf Problem}}
\newcommand\ca[1]{{\cal{#1}}}
\newcommand\lo[1]{_{\nano{#1}}}
\newcommand\hi[1]{^{\nano{#1}}}
\def\fsave{\mathrm{f\mbox{-}SAVE}}
\def\circtauyt{\topcirc \tau \lo Y \phantom{A}\hspace{-.17in}\trans }
\def\diag{{\mathrm{diag}}}
\renewcommand{\thesubsubsection}{\arabic{subsection}.\arabic{subsubsection}}
\def\cid{\stackrel{\mbox{\tiny $\cal D$}}\longrightarrow}
\def\argmin{\mathrm{argmin}}
\def\Range{\mathrm{Range}}
\def\Span{\mathrm{Span}}
\def\proof{\noindent {\sc Proof. }}
\def\Sigmaxx{\Sigma_{\mbox{\tiny $X\!X$}}}
\def\Sigmaxy{\Sigma_{\mbox{\tiny $X\!Y$}}}
\def\Sigmayx{\Sigma_{\mbox{\tiny $Y\!X$}}}
\def\Sigmayh{\Sigma_{\mbox{\tiny $Y\!H$}}}
\def\Sigmahh{\Sigma_{\mbox{\tiny $H\!H$}}}
\def\Sigmahy{\Sigma_{\mbox{\tiny $H\!Y$}}}
\def\Sigmayy{\Sigma_{\mbox{\tiny $Y\!Y$}}}
\def\cip{\stackrel{\mbox{\tiny $P$}}\rightarrow}
\def\As{{\mbox{\tiny $A$}}}
\def\Ss{{\mbox{\tiny $S$}}}
\def\Es{{\mbox{\tiny $E$}}}
\def\E{\mathbb{E}}
\def\R{\mathbb R}
\def\S{\mathbb S}
\def\F{{\cal F}}
\def\G{{\cal G}}
\def\H{{\cal{H}}}
\def\HT{\H \lo T}
\def\HX{\H \lo X}
\def\HY{\H \lo Y}
\def\L{{\cal L}}
\def\M{\mathfrak M}
\def\N{{\mathfrak {N}}}
\def\S{{\mathfrak{S}}}
\def\PCA{{\mbox{\tiny $\mathrm{PC}$}}}
\def\RR{{\mbox{\tiny $\mathrm{RR}$}}}
\def\given{| \,}
\def\central{{\mathfrak{S}_{\nano Y|X}}}
\def\cfield{{\G_{\nano Y|X}}}
\def\tsum{\textstyle{\sum}}
\def\Xb{\boldsymbol X}
\def\trans{^{\mbox{\tiny{\sf  T}}}}
\def\Yb{\boldsymbol Y}
\def\eqdis{\stackrel{\mbox{\tiny $\cal D$}}{=}}
\def\half{^{\mbox{\tiny $\frac{1}{2}$}}}
\def\mhalf{^{\mbox{\tiny $-\frac{1}{2}$}}}
\def\inv{^{\mbox{\tiny $-1$}}}
\def\thetabfs{{\mbox{\tiny $\thetabf$}}}
\def\Xbfs{{\mbox{\tiny $\Xbf$}}}
\def\var{\mathrm{var}}
\def\cov{\mathrm{cov}}

\def\subshat{\, \, \hat {\!\!\stens S}}
\def\shat{\ \hat{\!\! \sten S}}
\def\stenhat#1{\ \, \hat{\!\!\!  #1}}
\def\tr{\mathrm{tr}}
\def\vec{\mathrm{vec}}
\def\mat{\mathrm{mat}}
\def\sign{\mathrm{sign}}
\def\eop{\hfill $\Box$ \\
}
\def\trans{^{\mbox{\tiny{\sf T}}}}
\def\Xs{{\mbox{\tiny $X$}}}
\def\cid{\stackrel{\mbox{\tiny $\cal D$}}\rightarrow}
\def\iff{\Leftrightarrow}
\def\gpc{{\cal G}}
\def\dy{D_{{\mbox{\tiny $Y$}}}}
\def\du{D_{{\mbox{\tiny $U$}}}}
\def\kappax{\kappa_{\mbox{\tiny $X$}}}
\def\kappat{\kappa_{\mbox{\tiny $T$}}}
\def\kappay{\kappa_{\mbox{\tiny $Y$}}}
\def\Xsmall{\mbox{\tiny $X$}}
\def\Ysmall{\mbox{\tiny $Y$}}
\def\yygx{{\mbox{\tiny $YY|X$}}}
\def\xxsub{{\mbox{\tiny $XX$}}}
\def\yysub{{\mbox{\tiny $YY$}}}
\def\xysub{{\mbox{\tiny $XY$}}}
\def\yxsub{{\mbox{\tiny $YX$}}}
\def\Psub{{\mbox{\tiny $P$}}}
\def\ali{&\,}
\def\card{{\mathrm{card}}}
\def\spc{{\cal S}}
\def\ran{\mathrm{ran}}
\def\XYsub{{\mbox{\tiny $XY$}}}
\def\Xsub{{\mbox{\tiny $X$}}}
\def\Ysub{{\mbox{\tiny $Y$}}}
\def\central{\sten S \lo {Y|X}}
\def\quo{^\circ}
\def\of{{\nano {\circ}}}
\def\nano{\scriptscriptstyle}
\def\scirc{{\mbox{\tiny $\circ$}}}
\def\coxx{\Sigma_{\nano XX}}
\def\coxy{\Sigma_{\nano XY}}
\def\coyx{\Sigma_{\nano YX}}
\def\coyy{\Sigma_{\nano YY}}
\def\half{^{\nano 1/2}}
\def\nhalf{^{\nano -1/2}}
\def\rxy{R_{\nano XY}}
\def\ryx{R_{\nano YX}}
\def\hx{{\H_{\nano X}}}
\def\hy{{\H_{\nano Y}}}
\def\ltwox{{L_2(P_{\nano X})}}
\def\ltwoy{{L_2(P_{\nano Y})}}
\def\mean{\overline}
\def\cltwox{\mathrm{cl}_\ltwox}
\def\chx{\mathrm{cl}_\hx}
\def\equal{&=\, }
\def\real{{\mathbb R}}
\def\ddo{{$\ddot{\mbox{o}}$}}
\def\ddu{{$\ddot{\mbox{u}}$}}
\def\complete{\mathfrak{C}_{\nano Y|X}}
\def\completeoc{\mathfrak{C}_{\nano Y|X}\oc}
\def\tint{\textstyle{\int}}
\def\ka{\kappa}
\def\cenclass{\mathfrak{S} \lo {Y|X}}
\newcommand\oop[4]{#1 \lo {{\uppercase{#2}}({\uppercase{#3}} \otimes {\uppercase{#4}})}}
\def\oc{^{\nano \perp}}
\def\X{\cal{X}}
\def\Y{\cal{Y}}
\def\L2T{L \lo 2 (T)}
\def\L2TX{L \lo 2 (T\lo X)}
\def\L2TX{L \lo 2 (T\lo Y)}
\def\tsum{\textstyle{\sum}}
\def\ali{&\,}
\def\spn{{\rm{span}}}
\def\ali{&\,}
\def\adj{\hi *}
\def\eod{\end{document}}
\def\mpinv{\hi {\dagger}}
\def\shortbar{\rule[-.0em]{.04em}{.15em}}
\def\bbar{\mbox{\raisebox{-.05em}{$\,\overset{\overset{\shortbar}{\shortbar}}{\shortbar}\,$}}}
\newcommand\remind[1]{{\color{red}{{#1}}}}
\newcommand\ask[1]{{\color{blue}{{#1}}}}

\newcommand{\lbing}{\;\, \rule[0em]{.13em}{1.5em}}

%\tikzset
%{
%    x=2cm,% default value is 1cm.
%    y=3cm,% default value is 1cm.
%}

\def\lbing{
\!\raisebox{-2.3pt}{ \tikz[x=2.5pt,y=2.6pt]{\draw[line width=0.4pt](-1+4.8,-1)--(0+4.8,-2);\draw[line width=0.4pt](-1+4.8,1)--(0+4.8,2);\draw[line width=0.4pt](-1+4.8,1-2)--(-1+4.8,0+1);}}\hspace{0pt}}
\def\rbing{\!\!
\raisebox{-2.3pt}{ \tikz[x=2.5pt,y=2.6pt]{\draw[line width=0.4pt](-1-0.8,-2)--(0-0.8,-1);\draw[line width=0.4pt](-1-0.8,2)--(0-0.8,1);\draw[line width=0.4pt](0-0.8,-1)--(0-0.8,1);}}}

\def\cran{{\overline{{\rm{ran}}}}}
\def\dom{\mathrm{dom}}

\def\topcirc#1{\overset{\nano \circ}{#1}}
\newcommand{\magenta}[1]{{\leavevmode\color{magenta}{#1}}}
\newcommand{\cyan}[1]{{\leavevmode\color{cyan}{#1}}}
\newcommand{\brown}[1]{{\leavevmode\color{Brown}{#1}}}
\newcommand{\olivegreen}[1]{{\leavevmode\color{OliveGreen}{#1}}}
\newcommand{\mahogany}[1]{{\leavevmode\color{Mahogany}{#1}}}
\newcommand{\navyblue}[1]{{\leavevmode\color{NavyBlue}{#1}}}
\newcommand{\mulberry}[1]{{\leavevmode\color{Mulberry}{#1}}}
\newcommand{\redorange}[1]{{\leavevmode\color{RedOrange}{#1}}}
\def\card{{\mathrm{card}}}
\newcommand{\skyblue}[1]{{\leavevmode\color{SkyBlue}{#1}}}
\newcommand{\salmon}[1]{{\leavevmode\color{Salmon}{#1}}}
\def\loo#1{_{\nano{\mathrm{\uppercase{#1}}}}}

%\Large
\newfont{\rsfsten}{rsfs10 scaled 1100}
\newfont{\rsfstena}{rsfs10 scaled 800}
%\newfont{\rsfsten}{rsfs10 scaled 1300}
%\newfont{\rsfstena}{rsfs10 scaled 800}
\newfont{\rsfstenb}{rsfs10 scaled 500}
\newcommand{\sten}[1]{\mbox{\rsfsten #1}\,}
\newcommand{\stens}[1]{\mbox{\rsfstena #1}\,}
\newcommand{\stent}[1]{\mbox{\rsfstenb #1}}

\def\iff{\Leftrightarrow}

\def\dom{{\mathrm{dom}}}
\def\imply{\Rightarrow}
\def\central{\sten S \lo {Y|X}}

\def\central{\sten S \lo {Y|X}}
\def\hatcentral{\stenhat {\sten S} \lo {Y|X}}

\spacingset{1.7} % line spacing
% \spacingset{2} % line spacing

% NOTE: To produce blinded version, replace "0" with "1" below.
\newcommand{\blind}{1}

\if1\blind
{
  \title{\Large Collapsing Categories for Regression with Mixed Predictors}
  \author{\normalsize Chaegeun Song, Zhong Zheng, Bing Li, and Lingzhou Xue\\
    \normalsize  Department of Statistics, The Pennsylvania State University}
  \maketitle
} \fi

\if0\blind
{
  \bigskip
  \bigskip
  \bigskip
  \begin{center}
    {\LARGE\bf Collapsing categories for   regression with mixed predictors}
\end{center}
  \medskip
} \fi

\bigskip
\begin{abstract}
Categorical predictors are omnipresent in everyday regression practice: in fact, most regression data involve some categorical predictors, and this tendency is increasing in modern applications with more complex structures and larger data sizes. However, including too many categories in a regression model would seriously hamper accuracy, as the information in the data is fragmented by the multitude of categories. In this paper, we introduce a systematic method to reduce the complexity of categorical predictors by adaptively collapsing categories in regressions, so as to enhance the performance of regression estimation. Our method is based on the {\em pairwise vector fused LASSO}, which automatically fuses the categories that bear a similar regression relation with the response. We develop our method under a wide class of regression models defined by a general loss function, which includes linear models and generalized linear models as special cases. We rigorously established the category collapsing consistency of our method, developed an Inexact Proximal Gradient Descent algorithm to implement it, and proved the feasibility and convergence of our algorithm. 
Through simulations and an application to Spotify music data, we demonstrate that our method can effectively reduce categorical complexity while improving prediction performance, making it a powerful tool for regression with mixed predictors.
\end{abstract}

\noindent%

\section{Introduction}\label{sec:intro}

Many regression problems encountered in practice involve a mix of continuous and categorical variables. For such regression problems, observations are grouped based on several categorical variables, with regression coefficients for the continuous variables varying across groups. In the medical application, for example, patients may be grouped by gender, age, and types of treatments received, whereas the relationship between the response, such as the recovery time,  and the continuous predictors,  such as dosage and hospital stay duration,  may differ from group to group. The situation is similar to the Analysis of Covariance (ANCOVA) with heterogeneity in regression coefficients across groups. A fundamental question is how to reduce the complexity of the categorical variables to enhance the overall estimation accuracy of the regression relation. While methods  for reducing the complexity of continuous variables have undergone intense development in the forms of, for example,   variable selection \citep{tibshirani1996regression,fan2001variable}  and sufficient dimension reduction \citep{li1991sliced, cook1994using,li2018sufficient},   and while there also exist substantial research on reducing complexity of categorical variables in regression problems without continuous predictors \citep{bondell2009simultaneous, gertheiss2010sparse, 
 stokell2021modelling}, 
the problem of reducing the complexity of categorical variables in the presence of continuous predicting variables has received less attention. In this paper, we develop a systematic methodology for reducing the complexity of the categorical variables in regressions with mixed predictors, which relies on adaptively collapsing categories or levels of categories that display similar regression relations.

\noindent 

The importance of reducing the complexity of categorical variables has been recognized and the existing works on this problem are mainly in two directions: the first is for regressions where all predictors are categorical, and the goal is to combine levels within categories that display similar regression effect; the second concerns regressions with categorical and continuous predictors and the goal is to fuse the intercepts between different categories.  Specifically,  in the first direction, \citet{gertheiss2010sparse} used the fused LASSO penalty to collapse categorical levels. 
In the context of Analysis of Variance (ANOVA), where a continuous response is modeled using categorical predictors, \citet{bondell2009simultaneous} proposed CAS-ANOVA, which constrains the pairwise differences of dummy-coded coefficients, while \citet{stokell2021modelling} introduced SCOPE, which applies a nonconvex penalty for this purpose. See also \citet{gertheiss2012regularization}, \citet{oelker2014regularization}, and  \citet{tutz2017modelling}. 
These methods primarily focus on ANOVA-style modeling and inference, aiming to identify similar categorical levels within categories and fuse them together.
In the second direction, \citet{ma2017concave} proposed a concave pairwise fusion penalty to identify homogeneous subgroups. They assumed the subject-specific intercepts to represent the heterogeneity and fused the intercepts to partition the sample into subgroups. \citet{liu2021capturing} adapted their method to the repeated measures data. \cite{ohishi2021coordinate} considered generalized fused LASSO to identify groups with equal group-intercepts.   See \citet{tutz2016regularized} for more extensive literature reviews.

Recent years have witnessed a growing body of work for uncovering structural homogeneity in regression models. \citet{ke2015homogeneity} introduced the homogeneity pursuit to identify subgroups of regression coefficients in high-dimensional settings. This idea has since been extended to single-index models for panel data \citep{lian2021homogeneity} and to quantile regression in network autoregressive models \citep{liu2024two}. In parallel, \citet{tang2016fused} proposed a fused LASSO approach for clustering regression coefficients across multiple datasets in integrative analysis. While these methods have proven effective in detecting latent homogeneity and clustering regression coefficients, existing approaches do not accommodate categorical predictors by collapsing their associated coefficient vectors through fusion.

In this paper, we study a general regression problem involving both continuous and categorical predictors. Our objective is to reduce the complexity introduced by categorical variables by encouraging categories (or levels) that exhibit similar regression effects to fuse together. To achieve this, we propose a specially designed pairwise vector fused LASSO (PVF-LASSO). Our work advances the existing literature in several important ways. First, we move beyond regression models with purely categorical predictors \citep{gertheiss2010sparse, bondell2009simultaneous, stokell2021modelling} by allowing a sub-vector of continuous predictors, and unlike earlier studies that fuse only intercepts \citep{ma2017concave, liu2021capturing, ohishi2021coordinate}, our method fuses the entire regression functions, providing a more comprehensive view of parameter homogeneity. Second, we design a novel penalty that enables the fusion of categorical levels across different categorical variables, thereby creating additional opportunities for complexity reduction, and our method is different from standard variable-selection methods such as the LASSO \citep{bondell2009simultaneous, gertheiss2010sparse}: rather than selecting or discarding an entire categorical variable, our method allows selective collapsing of levels within a category, yielding a more refined simplification. In addition, we incorporate the adaptive LASSO \citep{zou2006adaptive} into our framework to develop the adaptive PVF-LASSO, which further enhances estimation and selection performance. We also formulate the method under a general nonlinear regression framework defined through a loss function, encompassing linear regression and generalized linear models as special cases. Finally, our method contrasts with the group LASSO \citep{yuan2006model} and grouping pursuit \citep{shen2010grouping}, where grouping structures are imposed on regression variables rather than on categorical levels across observations.

We develop both the theoretical and computational foundations of our methodology. On the theoretical side, we established category collapsing consistency, which means that the estimated groups of the categories coincide with the true groups with probability tending to 1. This is achieved by extending the irrepresentable condition \citep{zhao2006model} to the current setting and deriving the subdifferential (Karush–Kuhn–Tucker) conditions \citep{karush1939minima}, \citealp{kuhn1951nonlinear})
for the new problem. On the computational side, we develop an Inexact Proximal Gradient (IPG) Algorithm that iteratively conducts proximal gradient descent steps. In our IPG, since the non-smooth penalty is non-separable for different category levels, we solve a convex optimization problem in each iteration by applying the Block Coordinate Descent (BCD) algorithm to solve the dual subproblem. Given that the exact minimizers for the subproblems are unavailable, motivated by \cite{bonettini2016variable,lee2019inexact,lee2020inexact, zheng2024new, zheng2024adaptive,zheng2025new}, we solve the subproblems inexactly with adaptive stopping conditions. A rigorous convergence guarantee is provided for our IPG. Our algorithm overcomes the computational difficulty discussed in Section 5 of \cite{tutz2017modelling} when using the pairwise vector fused non-smooth penalty.

The rest of the paper is organized as follows.
In Section \ref{sec:lin}, we introduce the PVF-LASSO for linear regression and establish its category collapsing consistency. 
In Section \ref{sec:glm}, we extend PVF-LASSO to a wide class of nonlinear regression problems defined by a general loss function. We call this generalization the GPVF-LASSO (with G standing for ``general''). 
Also in this section, we introduce the adaptive version of the GPVF-LASSO for the general regression problem. 
In Section \ref{sec:opt}, we develop an algorithm to solve our non-smooth convex optimization problem, and rigorously establish the feasibility and convergence of our algorithms. 
In Section \ref{sec:sim}, we evaluate the performance of our methods by comparing them with several existing methods through simulation experiments. Section \ref{sec:spotify} applies our methods to a Spotify music dataset.
The supplementary material includes additional proofs.

\def\hi#1{^{#1}}

\section{Pairwise vector fused LASSO for linear regression}\label{sec:lin}

\subsection{The procedure}
Consider a regression problem with a vector-valued continuous predictor, $X \in \real \hi p $,  and a categorical predictor,  $U$, which takes values in the finite set $\{1, \ldots, m \}$. Our regression model can be written in matrix form as follows: conditioning on $U = u$, 
\begin{equation*}\label{eq:reg}
Y_u = X_u \beta_u + \epsilon_u,  \quad u = 1, \ldots, m,    
\end{equation*}
where $Y_u$ is $n_u \times 1$ vector, $X_u$ is $n_u \times p$ matrix, $\beta_u$ is $p \times 1$ coefficient vector, and $\epsilon_u$ is $n_u \times 1$ error vector with mean $0$ and variance $\sigma^2$.

\def\lo#1{_{#1}}

Although we have assumed $U$ to be a single categorical predictor, it actually also accommodates multi-category predictors.   This is because any set of categorical predictors can be combined into a single categorical predictor through a simple index mapping. Specifically, suppose we have $ k $ categorical predictors $ (U_1, \dots, U_k) $, where each $ U_i $ taking values in $\{1, \ldots, m \lo i\}$. We define a new categorical variable taking values in  the Cartesian product 
\begin{align*}
    \{1, \ldots, m \lo 1\} \times \cdots \times \{ 1, \ldots, m \lo k \}. 
\end{align*}
This transformation preserves all information from the original predictors while allowing the problem to be treated as a single categorical predictor case. More importantly, it enables the collapsing of category levels across categorical predictors. For this reason, and without loss of generality, we focus on a single categorical predictor $ U $ with levels $ \{1, \dots, m\} $.

We assume that, among $\{ \beta \lo 1, \ldots, \beta \lo m \}$, only $s$ ($s \le m$)  of them are distinct, and represent the distinct $\beta$'s as $\{ \gamma \lo 1, \ldots, \gamma \lo s \}$. We do not know which $\beta$'s are distinct or the number $s$ of distinct groups. Our goal is to collapse the identical $\beta$ by means of a sparse penalty that encourages the similar ones to fuse together. This leads to the following objective function 
\begin{equation}\label{eq:obj}
    L(\beta) = L(\beta_1, \ldots, \beta_m) = \sum_{u=1}^m  \|Y_u-X_u\beta_u\|^2 +\lambda\sum_{u<u^{\prime}}\|\beta_u-\beta_{u^{\prime}}\|,
\end{equation}
We call this procedure the {\em pairwise vector fused LASSO}, or PVF-LASSO. Throughout the remainder of the paper, we define $\| \cdot \|$ as the standard $\ell_2$ norm without further mentioning.

As we will see in the data application, sometimes  it is beneficial to introduce weights to the objective function (\ref{eq:obj}), like this: 
\begin{align}
  \label{eq:obj weighted}
    L(\beta) = L(\beta_1, \ldots, \beta_m) = \sum_{u=1}^m w \lo u \|Y_u-X_u\beta_u\|^2 +\lambda\sum_{u<u^{\prime}}\|\beta_u-\beta_{u^{\prime}}\|,
\end{align}
where $w \lo 1, \ldots, w \lo m$ are positive constants might depend on the category sample size $n \lo u$. Since this modification would not change the subsequent theoretical development in any significant way, we will keep the matter simple by taking $w \lo u =1$.

\def\ali{& \, }

\subsection{Definition of category collapsing consistency}
Let $\beta \lo 0$ be the true parameter value for $\beta$. 
Our goal in the rest of this section is to prove that, if $\hat \beta$ is the minimizer of $L (\beta)$, then, with probability tending to 1, 
\begin{align}\label{eq:category collapsing consistency: original}
\hat \beta \lo i = \hat \beta \lo j \ \mbox{if and only if} \ \beta \lo {0i} = \beta \lo {0j}, 
\end{align}
Let $C=\{(i,j): 0 \le i< j \le m \}$. 
Decompose $C$ as $C \lo 0 \cup C \lo 1$, where  $C \lo 0 =    \{ (i,j) \in C: \beta \lo {0i} \ne \beta \lo {0j} \}$, and $C \lo 1 = \{ (i,j) \in C: \beta \lo {0i} = \beta \lo {0j} \}$. Then, statement (\ref{eq:category collapsing consistency: original}) is equivalent to  
\begin{align*}
    \hat \beta \lo i \ne  \ali \hat \beta \lo j \mbox{ for all $(i,j) \in C \lo 0$}, \quad 
    \hat \beta \lo i =    \hat \beta \lo j \mbox{ for all $(i,j) \in C \lo 1$}. 
\end{align*}
If this event has a probability tending to 1, our method has  {\em category collapsing consistency}.

\def\hii#1{\hi{(#1)}}
\def\trans{\hi {\mathsf T}}

It will be more convenient to express this statement in terms of the distinct gradient vectors $\{\gamma \lo {01}, \ldots, \gamma \lo {0s}\}$, as defined earlier.  To do so, we first introduce a matrix that maps distinct gradient vectors $\{\gamma \lo 1, \ldots, \gamma \lo s \}$ to $\{\beta \lo 1, \ldots, \beta \lo m \}$, which may contain identical vectors. The 
set $\{1, \ldots, m \}$  is the union of $s$ disjoint sets, say $V \lo 1, \ldots, V \lo s$, with $\{\beta \lo i: i \in V \lo u \}$ being the same vector.   If the $m$ categorical levels are collapsible, then at least one $V \lo u$ is not a singleton. 
Without loss of generality, assume that $V \lo 1, \ldots, V \lo {s \lo 0}$ are not singletons, and $V \lo {s \lo 0 + 1}, \ldots, V \lo s$ are singletons. 
Let $r \lo 1, \ldots, r \lo s$ denote the cardinalities of $V \lo 1, \ldots, V \lo s$. Then  $r \lo 1 + \cdots + r \lo s = m$, and 
\begin{align*}
    V \lo 1 = \ali     \{1, \ldots, r \lo 1 \}, \ 
    V \lo 2 =   \{r \lo 1+1, \ldots, r \lo 1 + r \lo 2\}, \cdots, \\
    V \lo {s \lo 0} = \ali  \{ r \lo 1 + \cdots + r \lo {s \lo 0 -1}  + 1, \ldots, r \lo 1 + \cdots + r \lo {s \lo 0} \},  \ 
    V \lo {s \lo 0 + 1} =  \{r \lo 1 + \cdots + r \lo {s \lo 0} + 1 \}, \ldots, V \lo s = \{m\}. 
\end{align*}
  For each $i = 1, \ldots, s$, let $I \lo p \hii {r \lo i} $ be the $r \lo i p \times p$ matrix   $(I \lo p, \ldots, I \lo p)\trans$, and let  
\begin{align}\label{eq:matrix A}
 A = \begin{pmatrix}
I_p^{(r_1)} &    & 0 \\
  & \ddots &  \\
0 &   & I \lo p \hii {r \lo s}
\end{pmatrix}.
\end{align}
This matrix transforms the distinct $\gamma$ to the collapsible $\beta$; that is, $\beta = A \gamma$.  Note that, in practice, the non-singleton groups need not be the first $s \lo 0$ groups,  as the above notation suggests. Nevertheless, we can always perform a suitable permutation to obtain the assumed arrangement. The next lemma expresses the condition (\ref{eq:category collapsing consistency: original}) in terms of the matrix $A$.

\begin{lemma}\label{lemma:A hat gamma} Condition (\ref{eq:category collapsing consistency: original}) is equivalent to 
\begin{align*}%\label{eq:two conditions vecfuse}
\hat \beta = A \hat \gamma, \quad \mbox{and} \quad   
        \hat \gamma \lo 1, \ldots, \hat \gamma \lo s \ \mbox{are distinct}. 
\end{align*} 
\end{lemma}

\def\sign{\mathrm{sign}}
\def\signeq{\overset{s}{=}}

\subsection{Irrepresentable condition}

Recall that the variable selection consistency of classical LASSO for linear regression hinges on a condition called the irrepresentable condition, which regulates the dependence among predictors. The same applies to our category collapsing consistency. In this subsection, we derive an irrepresentable condition for category collapsing consistency. 

\def\hii#1{\hi{(#1)}}
\def\trans{\hi {\mathsf T}}

For  $i = 1, \ldots, m$, let  $E \lo i$ denote the $mp \times p$ matrix with its $i$th block being the identity matrix $I_p$ and all the other blocks being  0 matrices, that is, $E \lo i = (0, \cdots, 0, I \lo p, 0, \cdots 0)\trans$ with $I \lo p$ taking the $i$th position. In this notation $\beta \lo i = E \lo i \trans  \beta$.
For $u \in \{1, \ldots, s \lo 0\}$, let $B \lo u$ be the submatrix
$\{ E \lo i - E \lo {i+1}: i \in V \lo u \setminus \{r \lo 1 + \cdots + r \lo u\} \}$. Then, let 
\begin{align}\label{eq:matrix B}
    B = ( B \lo 1, \ldots, B \lo {s \lo 0} ). 
\end{align}
This matrix is of dimension $mp \times (r \lo 1 -1 + \cdots + r \lo {s \lo 0} - 1 ) p $. Since 
\begin{align*}
    r \lo 1 + \cdots + r \lo {s \lo 0} + \underbrace{1 + \cdots + 1}_{s - s \lo 0} = m, 
\end{align*}
we have 
$r \lo 1 + \cdots + r \lo {s \lo 0} - s \lo 0 = m - s. $
Thus $B$ is of dimension $mp \times (m-s) p$. 

For two sets  $S \lo 1$ and $S \lo 2$ in the same Euclidean space, let $S \lo 1 + S \lo 2$ denote the set $\{x \lo 1 + x \lo 2: x \lo 1 \in S \lo 1, x \lo 2 \in S \lo 2 \}$. For a matrix $M \in \real \hi {s\times t}$ and a set $S \subseteq \real \hi t$, let  $M S$ denote the set $\{ M x: x \in S \}$. The next three lemmas give some properties of the matrices $A$ and $B$, which are crucial for the subsequent developments. 
Let  $B \lo p (a,b)$ denote  the closed  ball in $\real \hi p$ centered at $a$ with radius $b$, that is, 
\begin{align*}
    B \lo p (a, b) = \{ x \in \real \hi p: \| x - a \| \le b \}. 
\end{align*}
Let $v \lo 0 \in \real \hi {mp}$ be defined as
\begin{align*}
  \sum \lo {(i,j) \in C \lo 0} \,  (E \lo i - E \lo j) (  \beta \lo {0i} -   \beta \lo {0j}) / \|  \beta \lo {0i} -   \beta \lo {0j} \|.  
\end{align*}
The term is, in fact, a part of the subdifferential of the penalty function corresponding to the distinct groups.   This construction will be further explained in Lemma \ref{lemma:subdifferential vecfuse}. 

Also, for an integer $k$, let $\mathbbm{1} \lo k$ be the $k$-dimensional column vector of 1's. For a set $S$ in the Euclidean space, let $\mathbbm{1} \lo k S$ be the $k$-dimensional column vector of sets $(S, \ldots, S)\trans $. 
The next assumption is the irrepresentable condition needed for our category collapsing consistency.

\def\inv{\hi {-1}}

\begin{assumption} \label{def:irrep}
\quad For each $u = 1, \ldots, s \lo 0$, 
   $ B \lo u  \trans X \trans X A ( A \trans X \trans X A ) \inv A \trans v \lo 0 \in  \mathbbm{1} \lo {r \lo u - 1}
        B \lo p (0, r \lo u )$. 
\end{assumption}
% \noindent 
This assumption is analogous to the irrepresentable condition for the LASSO \citep{zhao2006model}, which ensures that inactive variables cannot be too correlated with the active ones. In our context, a pair of categories being ``active'' means they are distinct, and being ``inactive'' means they are identical. Since the matrix $A$ is associated with distinct categories, and $B \lo u$'s are associated with identical categories, Assumption \ref{def:irrep} controls the correlation between the active pairs and inactive pairs.

\subsection{Establishing category collapsing consistency}

Let 
\begin{align}\label{eq:X Y beta}
    X = 
    \begin{pmatrix}
        X \lo 1 & & 0 \\
        & \ddots & \\
        0& & X \lo m 
    \end{pmatrix}, \quad Y = 
    \begin{pmatrix}
        Y \lo 1 \\
        \vdots \\
        Y \lo m 
    \end{pmatrix}, \quad 
    \beta = 
    \begin{pmatrix} 
    \beta \lo 1 \\
    \vdots \\
    \beta \lo m 
    \end{pmatrix}.
\end{align}
Then (\ref{eq:obj}) can be rewritten as 
\begin{align*}
    L \lo 1 ( \beta) = \| Y - X \beta \| \hi 2, \quad L \lo 2 (\beta ) = \sum \lo {(i,j) \in C} \|(E \lo i - E \lo j ) \trans \beta \|.
\end{align*}
Since the function $L(\beta)$ is not everywhere differentiable, we need to use its subdifferential to characterize its minimizer. 
The following proposition,  commonly known as the KKT theorem, characterizes the minimizer of a potentially nondifferentiable convex function. 
\begin{proposition}\label{proposition:kkt}
    Suppose $f: A \to \real$ is a convex function. Then the following statements are equivalent: 
    \begin{enumerate}
        \item $\hat x$ is the minimizer of $f$ over $A$; 
        \item $0 \in \partial f (\hat x)$, where $\partial f (\hat x)$ is the subdifferential at $\hat x$, defined as the set of vectors $\{ v: f(x) - f (\hat x) \ge v\trans (x - \hat x) \}$. 
    \end{enumerate}
\end{proposition}
\noindent 
To apply this result to our objective function (\ref{eq:obj}), we need   the subdifferential of the functions 
\begin{align*}
    f \lo {ij} (\beta) = \| \beta \lo i - \beta \lo j \| =  \|(E \lo i - E \lo j) \hi \top \beta \| 
\end{align*}
for any $i < j$. The next lemma gives the subdifferential of these functions. 

\begin{lemma}\label{lemma:subdifferential vecfuse} For any $\beta \lo 0 = \{\beta \lo {01}\trans, \ldots, \beta \lo {0m} \trans ) \trans  \in \real \hi {mp}$, 
   \begin{align}\label{eq:subdifferential}
       \partial f \lo {ij}   ( \beta \lo 0) = 
       \begin{cases}
       (E \lo i - E \lo j) (\beta \lo {0i} - \beta \lo {0j} ) / \| \beta \lo {0i} - \beta \lo {0j} \|  & \mbox{if} \ \beta \lo {0i}  \ne \beta \lo {0j} \\
           \{   ( E \lo i - E \lo j ) \alpha : \| \alpha \| \le 1 \} & \mbox{if} \ \beta \lo {0i}  = \beta \lo {0j}. 
       \end{cases}
   \end{align}
\end{lemma}

\def\spn{\mathrm{span}}

Lemma \ref{lemma:subdifferential vecfuse} characterizes the subdifferential of the vector fusing penalty function $f \lo {ij}(\beta) = \| \beta \lo i - \beta \lo j \|$ at any $\beta \lo 0$. When ${\beta}_{0i} \neq {\beta}_{0j}$, the penalty function is differentiable and the subgradient is the vector given by the first line in (\ref{eq:subdifferential}).  
When $\hat{\beta}_{0i} = \hat{\beta}_{0j}$, the penalty function is nondifferentiable, and the subdifferential becomes a set in the second line in (\ref{eq:subdifferential}). 

We now present several lemmas about matrices $A$ and $B$, which are important to the further development of our theory and method.

\begin{lemma}\label{lemma:remove sets vecfuse} If $A$ and $H \lo {ij}$ are as defined above, then 
\begin{align*}
    A \trans \left( \sum \lo {(i,j) \in C \lo 1} H \lo {ij}   \right) = \{0\}. 
\end{align*}
\end{lemma}

Lemma \ref{lemma:remove sets vecfuse} means that the subdifferentials corresponding to the collapsed pairs are orthogonal to the column space of $A$.
\begin{lemma}\label{lemma:orthogonal vecfuse}
    If $A$ and $B$ are as defined in (\ref{eq:matrix A}) and (\ref{eq:matrix B}), respectively, then $A \trans B = 0$. 
\end{lemma}

Lemma \ref{lemma:orthogonal vecfuse} shows the orthogonality of matrices $A$ and $B$. Since the matrix $A$ characterizes the distinct group means, and the matrix $B$ represents the within-group contrasts, this lemma shows that the linear subspaces representing the distinct group means and within-group contrasts are orthogonal. 
This orthogonality is key to our theoretical guarantees.
The above choice of $B$ is not crucial: our proof of the main theorem can be done using any $B \in \real \hi {p-q}$ such that $\spn (B) = \spn (A) \hi \perp$. 

\begin{lemma}\label{lemma:remove vector vecfuse} For any $k < \ell$, $k, \ell = 1, \ldots, s$, we have 
\begin{align*}
    \spn \left( \sum \lo {(i,j) \in V \lo k \times V \lo \ell} (E \lo i - E \lo j) \right) \subseteq \spn (A). 
\end{align*}
\end{lemma}

The next lemma asserts that the sum of $k$ balls centered at 0 is still a ball centered at 0 with its radius being the sum of the radius of the original $k$ balls.

\begin{lemma}\label{lemma:balls} Let $r \lo 1 > 0$, \ldots, $r \lo k > 0$ be positive constants, then 
\begin{align*}
    B \lo p (0, r \lo 1) + \cdots + B \lo p (0, r \lo k) = B \lo p (0,  r \lo 1 + \cdots r \lo k). 
\end{align*}
\end{lemma}

We are now ready to establish the category collapsing consistency. 

\begin{theorem}\label{theorem:category collapsing: linear case} 
If Assumption \ref{def:irrep} is satisfied and 
$\sqrt n \prec \lambda \prec n$,
then 
\begin{align*}%\label{eq:target of consistency}
    P \left( \hat \beta \lo i = \hat \beta \lo j \mbox{  \ if and only if \ } \beta \lo {0i} = \beta \lo {0j} \mbox{ for all $(i,j) \in C $}\right) \to 1. 
\end{align*}
\end{theorem}

Theorem~\ref{theorem:category collapsing: linear case} establishes the category collapsing consistency of the proposed estimator in the linear regression setting. Under the generalized irrepresentable condition in Assumption~\ref{def:irrep} and an appropriate choice of the tuning parameter $\lambda$, the theorem guarantees that the estimated collapse structure recovers the true underlying category collapse with probability tending to one. In other words, any pair of category level is collapsed into the same group if and only if they truly belong to the same group. 

\def\loo#1{\lo{(#1)}}

\def\of{\circ}

\section{Category collapsing  for general   regression}\label{sec:glm}

In this section, we go beyond the linear regression model and consider the category collapsing problem for a wide class of general nonlinear regression involving continuous and categorical predictors. We first lay out the framework and conditions for the general regression, and then consider in detail the special case of Generalized Linear Models (GLM;
\citealt{mccullagh2019generalized}).

\subsection{General regression}

As in the linear model, our data $X$, $Y$, and $\beta$ are defined as in (\ref{eq:X Y beta}). Given a general loss function 
$\rho   : \real \hi n   \times \real \hi n \to \real$, our nonlinear regression minimizes the following loss function: 
\begin{align*}
    L (\beta ) = \rho ( Y, X \beta) + \lambda \sum \lo {(i,j) \in C} \| \beta \lo i - \beta \lo j \|. 
\end{align*}
In the linear case, $\rho (Y, X \beta)$ takes the form $\| Y - X \beta \| \hi 2$. We write the linear predictor $X \beta$ as $\eta$ and use $\nabla \lo \eta \rho $ to denote the derivative  $\partial \rho  (Y, \eta) / \partial \eta$, which is a vector in $\real \hi n$. 

The following theorem establishes the category collapsing consistency for the general regression setting described above. Some of the regularity conditions required may not be immediately intuitive; however, their rationale and practical interpretation will be clarified in the next subsection through the special case of generalized linear models.

\begin{theorem}\label{theorem:category collapsing: nonlinear case} 
Suppose
\begin{enumerate}
    \item $\sqrt n \prec \lambda \prec n$; \vspace{-.1in}
    \item $L(\beta)$ has a unique minimizer; \vspace{-.1in}
    \item $\rho (y, \eta)$ is differentiable with respect to $\eta$; \vspace{-.1in}
    \item the solution $\hat \gamma$ to the equation 
    \begin{align*}
    %\label{eq:equation for hat gamma}
- A \trans  X \trans \, \nabla \lo \eta \, \rho  (Y, X A \gamma) 
    =  {\lambda}  A \trans \sum \lo {1 \le i < j \le s} \,  (E \lo i - E \lo j) (\gamma \lo i - \gamma \lo j) / \|  \gamma \lo i -  \gamma \lo j \|  
\end{align*}
is consistent. \vspace{-.4in} \\
\item $B \trans X \trans   Q \lo {XA}  \, \nabla \lo \eta \, \rho  (Y, XA   \gamma \lo 0) $ is of the order $O \lo P (n \hi {1/2})$; \vspace{-.1in}
\item the irrepresentable condition (Assumption \ref{def:irrep}) is satisfied. 
\end{enumerate}
Then 
\begin{align*}%\label{eq:target of consistency}
    P \left( \hat \beta \lo i = \hat \beta \lo j \mbox{  \ if and only if \ } \beta \lo {0i} = \beta \lo {0j} \mbox{ for all $(i,j) \in C $}\right) \to 1. 
\end{align*}
\end{theorem}

In Theorem \ref{theorem:category collapsing: nonlinear case}, the first and last conditions are the same as the linear case in Theorem \ref{theorem:category collapsing: linear case}. The second and third conditions are standard regularity assumptions: uniqueness of the minimizer ensures identifiability, while differentiability of the loss guarantees that the gradient is well-defined. The fourth condition is a technical condition derived from the KKT conditions for the fused estimator in general regression. If $\hat{\gamma}$ is not consistent, the left-hand side of the equation cannot vanish asymptotically, implying that no valid solution to the condition exists.
The fifth condition is also quite mild because the left-hand side resembles a score function (i.e., the derivative of the sum of $n$ log likelihood terms), which is usually of the order $O \lo P ( n \hi {1/2})$.  These conditions will be further discussed and justified under the generalized linear model in the next subsection. 

\subsection{Generalized linear models}

To give concrete intuitions about the category collapsing problem in general regression, in this subsection, we focus on an important special case: the generalized linear model. We use this model to illustrate the various quantities in the last subsection and explain why the conditions assumed in Theorem \ref{theorem:category collapsing: nonlinear case} are reasonable.    As before, let
\begin{align*}
 \{   (X \lo {ui}, Y \lo {ui}): u = 1, \ldots, m, i = 1, \ldots, n \lo u \}
\end{align*} 
be predictors in $\real \hi p$ and real-valued response. For each $u = 1, \ldots, m$, we assume that $Y \lo {ui} | X \lo {ui}$ has the exponential family distribution 
\begin{align*}
    f (y \lo {ui} | x \lo {ui}; \beta \lo u ) = \prod \lo {u = 1} \hi m \prod \lo {i=1} \hi {n \lo u} \exp \{ y \lo {ui} \theta  (\beta \lo u \trans x \lo {ui} ))  - b ( \theta  ( x \lo {ui} \trans \beta \lo u ) ) \}, 
\end{align*}
where $b: \real \to \real $ is a one-to-one convex function called the cumulant generating function uniquely associated with a specific exponential family, 
and $\theta: \real \to \real$ 
is a one-to-one function called the canonical parameterization function. 

Our loss function $\rho$ in this case is the negative log likelihood function 
\begin{align*}
\rho (Y, X \beta ) =     - \sum \lo {u=1} \hi m \sum \lo {i=1} \hi {n \lo u}[ Y \lo {ui}   \theta   (\beta \lo u \trans X \lo {ui} ) - b ( \theta  ( \beta \lo u \trans  X \lo {ui} ) )]. 
\end{align*}
Note that the right-hand side is indeed a function of $Y$ and $X \beta$, justifying the notation $\rho (Y, X \beta)$ on the left. 
The gradient function $\nabla \lo \eta \rho $ is derived straightforward  differentiation:
\begin{align*}
       (\partial  / \partial \eta \lo {ui} ) \rho ( Y, \eta ) =    -   \dot  \theta   (\eta \lo {ui} )  [ Y \lo {ui}   - \dot b ( \theta  (\eta \lo {ui} ) )  ], \quad u = 1, \ldots, m, \ i = 1, \ldots, n \lo u.  
\end{align*}
So, if we adopt the vector notation 
\begin{align*}
\ali    \eta =   X \beta, \quad 
   \dot \theta (\eta) =   \{ \dot \theta (\eta \lo {ui}): u = 1, \ldots, m, i = 1, \ldots, u \lo u \}, \\
\ali  \hspace{.2in}  \dot b ( \theta (\eta) ) =  \{ \dot b ( \theta (\eta \lo {ui} )): u = 1, \ldots, m, i = 1, \ldots, n \lo u \},  
\end{align*}
then our gradient function is 
\begin{align*}
    \nabla \lo \eta \, \rho (Y, X \beta  ) = \mathrm{diag} ( \dot \theta  (X \beta )) [ Y - ( \dot b \of \theta ) (X \beta) ]. 
\end{align*}

Next, let us justify the conditions in Theorem \ref{theorem:category collapsing: nonlinear case} through the generalized linear model.  Conditions 1 and 6 were also made in the linear case (Theorem \ref{theorem:category collapsing: linear case}), which have already been justified. The differentiability condition 3 is regarded as quite mild and is satisfied by all generalized linear models. Condition 2 has already been justified earlier in this subsection. So it remains to justify conditions 4 and 5. First, let us consider condition 4. 
In the case of generalized linear models, the relation in condition 4 specializes to 
\begin{align}\label{eq:right is small}
\begin{split}
 \ali   - n \inv  A \trans  X \trans \, \mathrm{diag} ( \theta  (X A \hat \gamma   )) [ Y - ( \dot b \of \theta ) (X A \hat \gamma) ] \\
 \ali \hspace{1in}   =  n \inv {\lambda}  A \trans \sum \lo {1 \le i < j \le s} \,  (E \lo i - E \lo j) (\hat \gamma \lo i - \hat \gamma \lo j) / \| \hat \gamma \lo i - \hat \gamma \lo j \|   
 \end{split}
\end{align}
Since $(\hat \gamma \lo i - \hat \gamma \lo j) / \| \hat \gamma \lo i - \hat \gamma \lo j \| $   has norm 1, the random vector $\sum \lo {1 \le i < j \le s} \,  (E \lo i - E \lo j) (\hat \gamma \lo i - \hat \gamma \lo j) / \| \hat \gamma \lo i - \hat \gamma \lo j \|$ is a bounded. Since $\lambda \prec n$, the right-hand side of the above equation is $o \lo P (1)$. For a generic $\gamma \in \real \hi s$, the left-hand side of (\ref{eq:right is small}) can be rewritten as 
\begin{align*}
\ali    - n \inv \sum \lo {u=1} \hi m \sum \lo {i=1} \hi {n \lo u} A \trans X \lo {ui} [ Y \lo {ui} - (\dot b \of \theta ) (X \lo {ui} A \gamma \lo u ) ].
\end{align*} 
When $\gamma  = \gamma \lo 0$, we have $E[ Y \lo {ui} - (\dot b \of \theta ) (X \lo {ui} A \gamma \lo u ) ] = 0$. So, by the Lindeberg central limit theorem, the quantity is of the order $O \lo p ( n \hi {-1/2})$. For any $\gamma \ne \gamma \lo 0$, we have $E[ Y \lo {ui} - (\dot b \of \theta ) (X \lo {ui} A \gamma \lo u ) ] \ne 0$, and so, by the weak law of large numbers, the above quantity converges in probability to a nonzero vector, which is not $o \lo P (1) $ as equation (\ref{eq:right is small}) requires. Thus, intuitively, any fixed $\gamma \ne \gamma \lo 0$ does not satisfy (\ref{eq:right is small}); only those asymptotically near $\gamma \lo 0$ would satisfy (\ref{eq:right is small}). Using this intuition, we can follow the argument in the proof of Theorem 8.1 in \cite{li2019graduate} to show that the solution to (\ref{eq:right is small}) is consistent. 

For the generalized linear model, the quantity in condition 5 can be rewritten as 
\begin{align*}
   \sum \lo {u=1} \hi m \sum \lo {i=1} \hi {n \lo u} B \trans X \lo {ui} [ Y \lo {ui} - (\dot b \of \theta ) (X \lo {ui} A \gamma \lo {0u} ) ] 
\end{align*}
Since $E [ Y \lo {ui} - (\dot b \of \theta ) (X \lo {ui} A \gamma \lo {0u} ) ]=0$, by the Lindeberg central limit theorem, the above quantity is of the order $O \lo P ( n \hi {1/2})$. Thus, condition 5 is verified.

\subsection{Adaptive pairwise vector fused LASSO}\label{sec:ada}

To enhance the performance, we adapt
 the adaptive LASSO \citep{zou2006adaptive} for category collapsing.  Due to limited space, we omit the theoretical development of category collapsing consistency and the oracle property.  Unlike the pairwise vector fused LASSO, which uniformly shrinks all pairwise differences toward zero, the adaptive version strengthens the penalty on the differences that are already small, while allowing larger differences to survive.

For the linear model, let $\hat \beta^{\mathrm{ols}}$ denote the ordinary least squares (OLS) estimate. Let
$$w_{uu^\prime} = \frac{1}{\|\hat \beta_u^{\mathrm{ols}}-\hat \beta_{u^{\prime}}^{\mathrm{ols}}\|^\gamma}$$ for some constant $\gamma > 0$. We define the  objective function for adaptive PVF-LASSO as
\begin{equation*}\label{eq:adaobj}
    L(\beta) = L(\beta_1, \ldots, \beta_m) = \sum_{u=1}^m \|Y_u-X_u\beta_u\|^2 +\lambda\sum_{u<u^{\prime}}w_{uu^\prime}\|\beta_u-\beta_{u^{\prime}}\|,
\end{equation*}
For the general regression model, let $\hat \beta \hi {\mathrm{unp}}$ be the unpenalized minimizer of the objective function $\rho ( Y, X \beta)$, and let 
$w_{uu^\prime} = \frac{1}{\|\beta_u^{\mathrm{unp}}-\beta_{u^{\prime}}^{\mathrm{unp}}\|^\gamma}. $ We define  the objective function for the adaptive GPVF-LASSO for general regression as  
\begin{equation*}\label{eq:adaglmobj}
    L(\beta)
    = \rho ( Y, X \beta)  + \lambda\sum_{u<u^{\prime}}w_{uu^\prime}\|\beta_u-\beta_{u^{\prime}}\|.
\end{equation*}

The adaptive pairwise fused vector penalty can be solved using the same efficient algorithm as the pairwise fused vector penalty. The details are provided in Section \ref{sec:opt}.

\section{Optimization}\label{sec:opt}

\subsection{Inexact proximal gradient algorithm for linear regression}

In this subsection, we develop an  inexact proximal gradient algorithm (IPG) for solving the optimization problem
\begin{equation}\label{eq_opt_problem_lr}
    \min_{\beta\in \mathbb{R}^{mp}} L(\beta) =  \sum_{u=1}^m \|Y_u-X_u\beta_u\|^2_2              +\sum_{u<u^{\prime}}\lambda_{uu^{\prime}} \, \|\beta_u-\beta_{u^{\prime}}\|
\end{equation}
where, for PVF LASSO, we set   $\lambda_{uu^\prime} = \lambda >0$, and for adaptive PVF, we set   $\lambda_{uu^\prime} = \lambda w_{uu'}$. Since  $L(\beta)$ is convex and is the sum of the smooth part $L_1(\beta) = \sum_{u=1}^m \|Y_u-X_u\beta_u\|^2_2$ and the non-smooth part $L_2(\beta) = \sum_{u<u^{\prime}}\lambda_{uu^{\prime}} \, \|\beta_u-\beta_{u^{\prime}}\|$, we iteratively update $\beta$ following the idea of the proximal gradient method (See Chapter 4.2 in \cite{parikh2014proximal}). At the $t$-th iteration where $\beta$ is initialized as $\beta^t = \{\beta_u^t\}_{u=1}^m$, for any $u\in [m]$, we denote 
\begin{align*}
  G_u^t = \ali  \frac{\partial L_1}{\partial \beta_u}(\beta^t) = 2X_u^\top(X_u\beta_u^t - Y_u), \\
  \alpha_u = \ali  \left(\left\|\frac{\partial^2 L_1}{\partial \beta_u\partial \beta_u^\top}(\beta^t)\right\|\right)^{-1} = \left(2\|X_u^\top X_u\|\right)^{-1}, \\
  \hat{\beta}_u^t = \ali  \beta_u^t - \alpha_u G_u^t,  
\end{align*}  
where we have used the notation $\alpha_u$ instead of $\alpha_u^t$ because, for linear regression,  the Hessian matrix ${\partial^2 L_1(\beta)}/{\partial \beta\partial \beta^\top}$ does not depend on $\beta$, and thus $\{\alpha_u\}_{u=1}^m$ need  not   be updated through the  iterations over $t$. The subproblem is given by 
\begin{equation}\label{subproblem_ls}
    \min_{\beta\in\mathbb{R}^{mp}} \mathcal{L}_t(\beta) := \sum_{u=1}^m\left(\frac{1}{2\alpha_u}\|\beta_u - \hat{\beta}_u^t\|^2_2\right)+\sum_{u<u^{\prime}}\lambda_{uu^\prime}\|\beta_u-\beta_{u^{\prime}}\|.
\end{equation}
The function $\mathcal{L}_t(\beta)$ is strongly convex with regard to $\beta$.  The update of $\beta$ is given by
\begin{equation}\label{inexact update}
    \beta^{t+1} \approx \argmin_{\beta\in\mathbb{R}^{mp}} \mathcal{L}_t(\beta).
\end{equation}
Here, ``$\approx$" means that the subproblem \eqref{subproblem_ls} is solved inexactly. Note that  \eqref{inexact update} is different from the standard proximal gradient algorithm using exact minimization. This is because the analytic solution for \eqref{subproblem_ls} is unavailable, and thus exact minimization is impossible in practice. Following the idea of \cite{bonettini2016variable,lee2019inexact,lee2020inexact,zheng2024new,zheng2024adaptive}, we solve the subproblem \eqref{subproblem_ls} inexactly such that $\beta^{t+1}$ satisfies
\begin{equation}\label{eq_cond}
    \mathcal{L}_t(\beta^{t+1}) - \min_{\beta\in\mathbb{R}^{mp}} \mathcal{L}_t(\beta) \leq \tau \, \left(\mathcal{L}_t(\beta^{t}) - \mathcal{L}_t(\beta^{t+1})\right).
\end{equation}
Here, $\tau >0$ is a hyperparameter. For all the numerical experiments in the paper, we set $\tau=0.1$. Condition \eqref{eq_cond} controls the accuracy of the solution of the subproblem,  guaranteeing a sufficient decrease of the objective function in each iteration and the overall convergence. In the meantime, it circumvents having to solve the subproblem \eqref{subproblem_ls} to an unnecessarily high accuracy. We call the above procedure the Inexact Proximal Gradient (IPG) and provide it in Algorithm \ref{alg:IPG-ls}. It is only a prototype algorithm because we have not specified the subproblem solver, which will be discussed in the next subsection.

 \begin{algorithm}[!ht]  
\caption{IPG --  A Prototype}
\label{alg:IPG-ls}
\begin{algorithmic}
\STATE \textbf{Input:} Initial point $\beta^0\in\mathbb{R}^{mp}$, $\tau>0$.
\FOR{$t = 0, 1, \ldots, $}
\STATE Find $\beta^{t+1}\in\mathbb{R}^{mp}$ by some subproblem solver such that \eqref{inexact update} holds.
\STATE Stop when some termination condition is triggered.
\ENDFOR
\end{algorithmic}
\end{algorithm}

Theorem \ref{thm_opt_ls} rigorously establishes the convergence of the IPG algorithm.

\begin{theorem}\label{thm_opt_ls}
    If $\Omega = \{\beta\in\mathbb{R}^{mp}:L(\beta)\leq L(\beta^0)\}$ is a bounded set, then any clustering point of the sequence $\{\beta^t\}_{t=0}^\infty$ is a minimizer of $L(\beta)$.
\end{theorem}

The boundness condition on $\Omega$ required in Theorem \ref{thm_opt_ls} holds when $L(\beta)$ is coercive, which is satisfied when $\mbox{rank}(X_u) = p$ for all $u \in [m]$.

For ease of the presentation, we expand the definition for $\{\lambda_{uu'}\}_{u<u'}$ to $\{\lambda_{uu'}\}_{u\neq u'}$ such that $\lambda_{uu'} = \lambda_{u'u}$ when $u>u'$.
The difficulty of solving \eqref{inexact update} lies in the non-separable nonsmooth part $L_2(\beta)$. Using the fact that $L_2(\beta) = \sum_{u<u^{\prime}} \lambda_{uu'}\|\beta_u-\beta_{u^{\prime}}\| =  \, \sum_{u\neq u^{\prime}}({\lambda_{uu'}}/{2})\|\beta_u-\beta_{u^{\prime}}\|$, we write the Lagrangian function as follows:
\begin{equation*}\label{eq_lag_ls}
    \mathcal{L}_t(\beta,v,q) = \sum_{u=1}^m\left(\frac{1}{2\alpha_u}\|\beta_u - \hat{\beta}_u^t\|^2\right)+\frac{\lambda_{uu'}}{2}\sum_{u\neq u^{\prime}}\|v_{uu^\prime}\| + \sum_{u\neq u^\prime} \langle q_{uu^\prime}, \beta_u-\beta_{u^{\prime}} - v_{uu^\prime}\rangle
\end{equation*}
where $q,v\in\mathbb{R}^{m(m-1)p}$ represent $\{q_{uu^\prime}\}_{u\neq u'}$ and $\{v_{uu^\prime}\}_{u\neq u'}$ respectively, and $q_{uu^\prime},v_{uu^\prime}\in\mathbb{R}^p$. Denote 
\begin{align*}
\mathcal{D}_t(q) = \ali \sum_{u=1}^m\left(\frac{\alpha_u}{2}\left\|\sum_{u^\prime:u^\prime\neq u} (q_{uu^\prime} - q_{u^\prime u})\right\|^2 - \left\langle \sum_{u^\prime:u^\prime\neq u} (q_{uu^\prime} - q_{u^\prime u}), \hat{\beta}_u^t\right\rangle\right), \\
Q_{u}(q) = \ali \sum_{u^\prime:u^\prime\neq u} (q_{uu^\prime} - q_{u^\prime u}), \quad \forall \ u\in [m].
\end{align*}
We know that 
$$\min_{\beta\in\mathbb{R}^{mp},v\in\mathbb{R}^{m(m-1)p}} \mathcal{L}_t(\beta,v,q) = -\mathcal{D}_t(q) - \infty \times \sum_{u\neq u^\prime} \mathbf{1}[\|q_{uu^\prime}\|>\lambda_{uu'}/2] $$
with the link function
$$\beta^{t+1}(q) = \{\beta_u^{t+1}(q)\}_{u=1}^m, \quad \beta_u^{t+1}(q) = \hat{\beta}_u^t - \alpha_u Q_u(q), \quad    \forall u\in [m].$$
Thus, we can solve the dual problem as follows:
\begin{equation}\label{eq_dual_ls}
    \min_{q\in\mathbb{R}^{m(m-1)p}} \mathcal{D}_t(q) \quad \mbox{subject to} \quad \|q_{uu'}\|\leq \lambda_{uu'}/2, \quad \forall \ u\neq u^\prime.
\end{equation}
In \eqref{eq_dual_ls}, the constraints are specified individually for each pair of indices $u\neq u^\prime$,
% the constraints are separated for different indices $uu^\prime$ \blue{(Bing: Please rephrase this: I don't understand what ``separate'' means)}, 
and the partial minimizers are computed as follows:
\begin{equation}\label{eq_cord_update_ls}
    \hat{q}_{uu^\prime}(q_{-uu^\prime}) := \mbox{Proj}_{\|\cdot\|\leq \lambda_{uu'}/2}\left(\frac{\alpha_{u^\prime}\left(Q_{u^\prime}(q) + q_{uu^\prime}\right) + \hat{\beta}_u^t - \alpha_u(Q_u(q) - q_{uu^\prime}) - \hat{\beta}_{u^\prime}^t}{\alpha_{u}+\alpha_{u^\prime}}\right), \quad \forall u\neq u^\prime,
\end{equation}
$$\hat{q}_{uu^\prime}(q_{-uu^\prime})\in \argmin_{q_{uu^\prime}\in \mathbb{R}^p:\|q_{uu^\prime}\|\leq \lambda_{uu'}/2} \mathcal{D}_t(q), \quad \forall u\neq u^\prime.$$
Here, $q_{-uu^\prime}$ represents all the components except $q_{uu^\prime}$; that is 
\begin{align*}
q \lo {-u u' } = \{q_{v v'}: (v, v') \in [m] \times [m], v \ne v', (v,v') \ne (u,u')\}.  
\end{align*}
In addition, for any $x\in\mathbb{R}^p$, we denote the projection to the set $\{x\in\mathbb{R}^p:\|x\|\leq \lambda_{uu'}/2\}$ as
$$\mbox{Proj}_{\|\cdot\|\leq \lambda_{uu'}/2}(x) = \begin{cases}
    x, & x\leq \lambda_{uu'}/2,\\
    \lambda_{uu'}x/(2\|x\|), & x> \lambda_{uu'}/2.
\end{cases}$$
Since neither $Q_{u^\prime}(q) + q_{uu^\prime}$ nor $Q_u(q) - q_{uu^\prime}$ depend on $q_{uu^\prime}$,  they are, in fact,  functions of $q_{-uu^\prime}$. Thus, we can use the block coordinate descent (BCD) algorithm \citep{tseng2001convergence} that iteratively updates $q$. In each iteration, it  further iteratively updates $q_{uu^\prime}$ by $\hat{q}_{uu^\prime}(q_{-uu^\prime})$ for $(u,u^\prime)\in \mathcal{I} := ([m]\times [m])\backslash \{(u,u):u\in [m]\}$.

Next, we propose a termination condition for BCD that guarantees \eqref{inexact update}. For any dual solution $q\in\mathbb{R}^{m(m-1)p}$ such that $\|q_{uu^\prime}\|\leq \lambda_{uu'}/2,\forall (u,u^\prime)\in\mathcal{I}$, by weak duality, we have
$$\mathcal{L}_t(\beta^{t+1}(q)) + \mathcal{D}_t(q)\geq \mathcal{L}_t(\beta^{t+1}(q)) - \min_{\beta\in\mathbb{R}^{mp}}\mathcal{L}_t(\beta).$$
Hence, 
\begin{equation}\label{eq_sufficient_stopping}
    \mathcal{L}_t(\beta^{t+1}(q)) + \mathcal{D}_t(q)\leq \tau(\mathcal{L}_t(\beta^t) - \mathcal{L}_t(\beta^{t+1}(q)))
\end{equation}
is a sufficient condition for \eqref{inexact update} with $\beta^{t+1} = \beta^{t+1}(q)$, and we can use it as the termination condition for BCD. Algorithm \ref{alg:IPG-ls-sub} summarizes the above procedure. 
\begin{algorithm}[ht]  
\caption{BCD for \eqref{subproblem_ls}}
\label{alg:IPG-ls-sub}
\begin{algorithmic}
\STATE \textbf{Input:} Initial point $q\in\mathbb{R}^{m(m-1)p}$ such that $\|q_{uu^\prime}\|\leq \lambda_{uu'}/2,\forall u\neq u'$, $\tau>0$.
\WHILE{\eqref{eq_sufficient_stopping} does not hold,}
\FOR{$(u,u^\prime)\in \mathcal{I}$}
\STATE $q_{uu^\prime}\leftarrow \hat{q}_{uu^\prime}(q_{-uu^\prime})$.
\ENDFOR
\ENDWHILE
\STATE Output: $\beta^{t+1} = \beta^{t+1}(q)$.
\end{algorithmic}
\end{algorithm}

\noindent 
Theorem \ref{thm_opt_ls_sub} shows the feasibility of Algorithm \ref{alg:IPG-ls-sub}.
\begin{theorem}\label{thm_opt_ls_sub}
    If $\beta^t\neq S_t(\beta^t)$, Algorithm \ref{alg:IPG-ls-sub} terminates with finite number of ``while" loops.
\end{theorem}

Note that $\beta^t\neq S_t(\beta^t)$  in Theorem \ref{thm_opt_ls_sub} is equivalent to $\beta^t\notin \argmin_{\beta\in\mathbb{R}^n} L(\beta)$.
In numerical experiments, we   terminate Algorithm \ref{alg:IPG-ls} when we find $\beta^{t+1}$ such that $\|\beta^t - \beta^{t+1}\|\leq 10^{-7}$.

\subsection{Inexact proximal gradient algorithm for general regression}

Next, we develop the inexact proximal gradient algorithm (IPG) for solving the optimization problem of general regression: 
\begin{equation}\label{eq_opt_problem_glm}
    \min_{\beta\in \mathcal{X}} L(\beta) =   \rho ( Y, X \beta ) +\sum_{u<u^{\prime}} \lambda_{uu^{\prime}}\|\beta_u-\beta_{u^{\prime}}\| := L \lo 1 (\beta) + L \lo 2 (\beta). 
\end{equation}
Here, $\beta$ stands for the vector  $( \beta \lo 1 \trans, \ldots, \beta \lo m \trans) \trans$, $\mathcal{X}\subseteq \mathbb{R}^{mp}$ is the feasible region for $\beta$, $\lambda_{uu^{\prime}}>0$ is a constant for any $u<u^{\prime}$, and  $L \lo 1 (\beta)$ is the general loss function $\rho ( Y, X \beta )$. We make the following assumptions for the setting of general regression.
\begin{assumption}\label{assu_glm_opt}
    $L(\beta)$ in \eqref{eq_opt_problem_glm} satisfies the following conditions.
    \begin{enumerate}
        \item $\mathcal{X}$ is nonempty, open, and convex. 
         \item The function $L_1(\cdot)$ is non-constant, convex and smooth on $\mathcal{X}$. Furthermore, there exists $\Tilde{\beta}\in \mathcal{X}$ such that $L(\Tilde{\beta}) =\min_{\beta\in\mathcal{X}} L(\beta)$. 
        \item For any $\hat{\beta}\in \mathcal{X}$, $\Omega(\hat{\beta}):=\{\beta\in\mathcal{X}: L(\beta)\leq L(\hat{\beta})\}$ is bounded, and $\nabla L_1(\beta)$ is Lipschitz continuous on $\Omega(\hat{\beta})$.
    \end{enumerate}
\end{assumption}
Under Assumption \ref{assu_glm_opt},   $L(\beta)$ is the sum of the smooth part $L_1(\beta)$ and the nonsmooth part $L_2(\beta)$. 
Similar to the linear case, at the $t$-th iteration, where $\beta$ is initialized as $\beta^t = \{\beta_u^t\}_{u=1}^m$, for any $u\in [m]$, we denote $G_u^t := \frac{\partial L_1}{\partial \beta_u}(\beta^t),\alpha_u >0$ as the step size, and $\hat{\beta}_u^t = \beta_u^t - \alpha_u G_u^t.$
Here, we use the same step size $\alpha \lo u$ for all iterations in $t$.  The subproblem is the same as \eqref{subproblem_ls} in the linear case. We still solve it inexactly as in \eqref{inexact update} such that \eqref{eq_cond} holds, and the IPG algorithm still works as in Algorithm \ref{alg:IPG-ls} except that we initialize $\beta^0$ in $\mathcal{X}$. Here, we remark that the subproblem \eqref{subproblem_ls} is still solved on $\mathbb{R}^{mp}$ so that the feasible region constraint is relaxed. Thus, the subproblem under the general regression takes the same form as that for the linear model. We can apply Algorithm \ref{alg:IPG-ls-sub} to solve it, and Theorem \ref{thm_opt_ls_sub} still holds.

Theorem \ref{thm_opt_glm} formally provides the convergence of IPG under the general regression.
\begin{theorem}\label{thm_opt_glm}
    For the optimization problem \eqref{eq_opt_problem_glm}, suppose that Assumption \ref{assu_glm_opt} holds, and $\{\beta^t\}_{t=0}^\infty$ is generated by Algorithm \ref{alg:IPG-ls} 
    with $\beta^0\in\mathcal{X}$. Then there exists an $\alpha'>0$ such that,  when $\alpha_u \leq \alpha'$ for all $u\in [m]$,  any cluster point of the sequence $\{\beta^t\}_{t=0}^\infty$ is a minimizer of $L(\beta)$ on $\mathcal{X}$.
\end{theorem}

To conclude this subsection, we illustrate the choices of the step sizes $\alpha_u,u\in [m]$ for the logistic model, where
$$
L(\beta) = \sum_{u=1}^m\left(\mathbf{1}^\top \log (1+ \exp(X_u\beta_u)) - \langle X_u^\top Y_u,\beta_u\rangle \right)+\sum_{u<u^{\prime}}\lambda_{uu^\prime}\|\beta_u-\beta_{u^{\prime}}\|.
$$
Here, $\exp(\cdot)$, $\log(\cdot)$, and other operations on scalars represent element-wise mappings when applied to vectors. We can find that $\frac{\partial L_1}{\partial \beta_u}(\beta) = X_u^\top(p_u(\beta_u) - Y_u)$ where $p_u(\beta_u) = \frac{1}{1+\exp(-X_u\beta_u)}\in\mathbb{R}^{n_u}$ and $\frac{\partial^2 L_1}{\partial \beta_u\partial \beta_u^\top}(\beta) = X_u^\top \mbox{diag}(p_u(\beta_u)\odot (1-p_u(\beta_u)))X_u$ where $\odot$ represents element-wise multiplication. Thus, we can let $\alpha_u = \left(\|X_u^\top X_u\|/4\right)^{-1}$ so that $\left\|\frac{\partial^2 L_1}{\partial \beta_u\partial \beta_u^\top}(\beta)\right\|\leq \alpha_u^{-1},\forall \beta\in\mathbb{R}^{mp}$. This ensures $\mathcal{L}(\beta;\beta') - \mathcal{L}(\beta';\beta')\geq L(\beta) - L(\beta'),\forall \beta,\beta'\in\mathbb{R}^{mp}$, which satisfies the requirement (S22) in supplementary material.

\section{Simulation}\label{sec:sim}

In this section, we conduct simulations to evaluate the performance of the proposed PVF-LASSO and adaptive PVF-LASSO in identifying the groups in the categories and estimating the distinct regression coefficient vectors. 

Our simulation studies contain a categorical variable with 6 categories  ($ u = 1, \ldots, 6$) and 3 continuous variables ($p = 3$).  Among the regression coefficient vectors for the 6 categories, 
the first pair is the same, the second pair is the same, and the last pair is the same, as follows: 
\begin{align*}
  \beta_1 = \beta_2 = 
\begin{pmatrix}
    1 \\
    1.3 \\
    -1.3
\end{pmatrix}, \quad 
\beta_3 = \beta_4 = 
\begin{pmatrix}
0.5 \\
-0.5 \\
0.5  
\end{pmatrix}, \quad 
 \beta_5 = \beta_6 = 
\begin{pmatrix}
    -1.5 \\
    -0.8 \\
    0.8
\end{pmatrix}.    
\end{align*}
We consider the following two regression scenarios. 
\begin{itemize}
    \item Scenario 1 (Linear regression). For each category $u = 1, \ldots, 6$, the data is generated from $Y \lo {ui} = X_{ui}\beta_u + \epsilon_{ui}$, $i = 1, \ldots, n \lo u$, where $\{X_{ui} \} \indep \{\epsilon \lo {ui} \}$, $\{X \lo {ui} \}$ and $\{\epsilon \lo {ui} \}$ are i.i.d. $N(0,\sigma^2)$ variables. We use two noise levels, $\sigma^2 = 1$ and $\sigma^2= 5$, to assess the robustness of the methods under different signal-to-noise ratios. The total sample size is $n= 300 $   with $n_u = 50$ observations per category.
    \item Scenario 2 (Logistic regression). The linear model is replaced by $g(\mu \lo {ui}) = X \lo {ui} \beta \lo u$, where $g$ is the logistic link function, $Y \lo {ui} \sim \mbox{Bernoulli}(\mu \lo {ui})$,  {and $X \lo {ui}\sim N(0,\sigma^2)$. The two variance settings are replaced by  $\sigma^2 = 1$ and $\sigma^2= 1/10$, which yield a reasonable range of signal-to-noise ratio in this regression setting. The total number of observations is $n = 600$ and $n \lo u = 100$ for each $u$}. 
\end{itemize}

To ensure a robust evaluation, we run
100 independent simulations for each scenario. In each simulation, we split the data for each category into training and testing sets, with 80\% allocated to the training set. We then estimate the coefficient vector $\beta_u$ using the training set and calculate the mean squared error (MSE) for the prediction on the test set to assess performance. 

We compare the proposed PVF-LASSO and adaptive PVF-LASSO with three other methods, respectively called Oracle, Single, and Separate. The Oracle, which is used as a benchmark for comparison, performs regression based on the true group, combining categories with identical $\beta_u$ values to estimate shared coefficients using maximum likelihood estimation. The Single assumes a single $\beta_u$ across all categories, ignoring the existence of categories. The Separate treats each category independently and estimates a separate $\beta_u$ for each, ignoring the existence of the groups. For the  PVF-LASSO and the adaptive PVF-LASSO, we use five-fold cross-validation to select the tuning parameter $\lambda$ and adaptively estimate the coefficients without requiring prior group knowledge. For the adaptive PVF-LASSO, $\gamma = 2$ and $\gamma = 0.5$ are picked for Scenarios  1 and 2, respectively.

\begin{figure}
    \centering
    \begin{minipage}{0.5\textwidth}
        \centering
        \includegraphics[width=\linewidth]{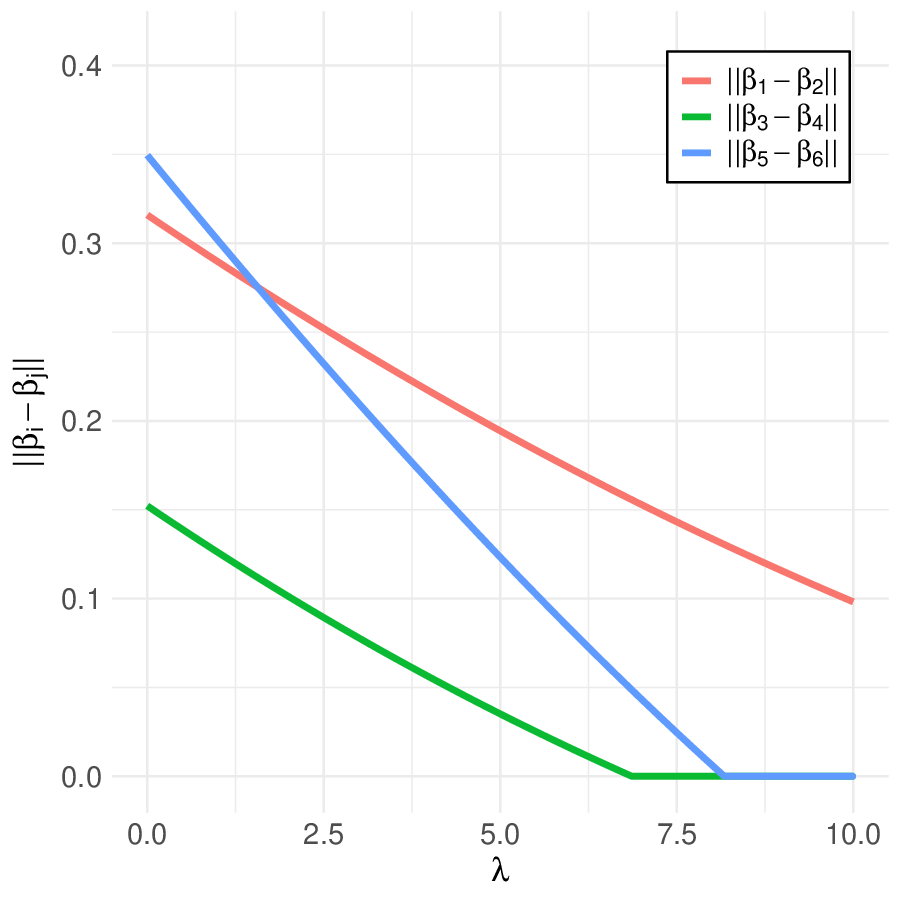}
    \end{minipage}\hfill
    \begin{minipage}{0.5\textwidth}
        \centering
        \includegraphics[width=\linewidth]{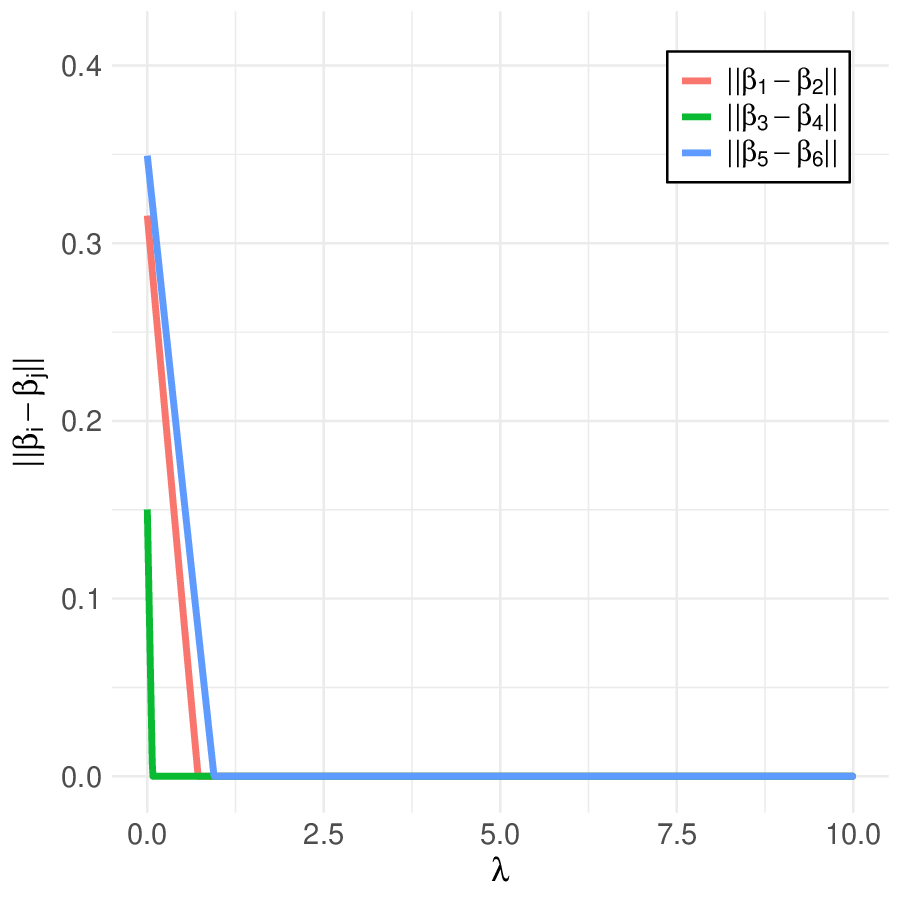}
    \end{minipage}
    \begin{minipage}{0.5\textwidth}
        \centering
        \includegraphics[width=\linewidth]{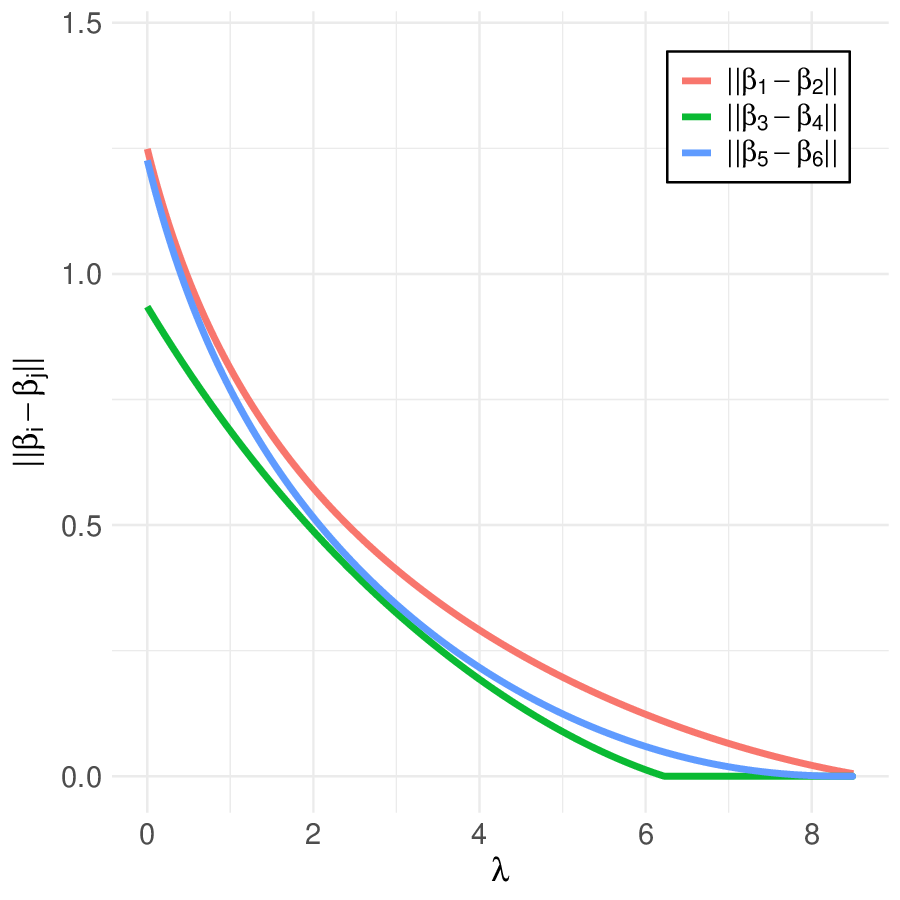}
    \end{minipage}\hfill
    \begin{minipage}{0.5\textwidth}
        \centering
        \includegraphics[width=\linewidth]{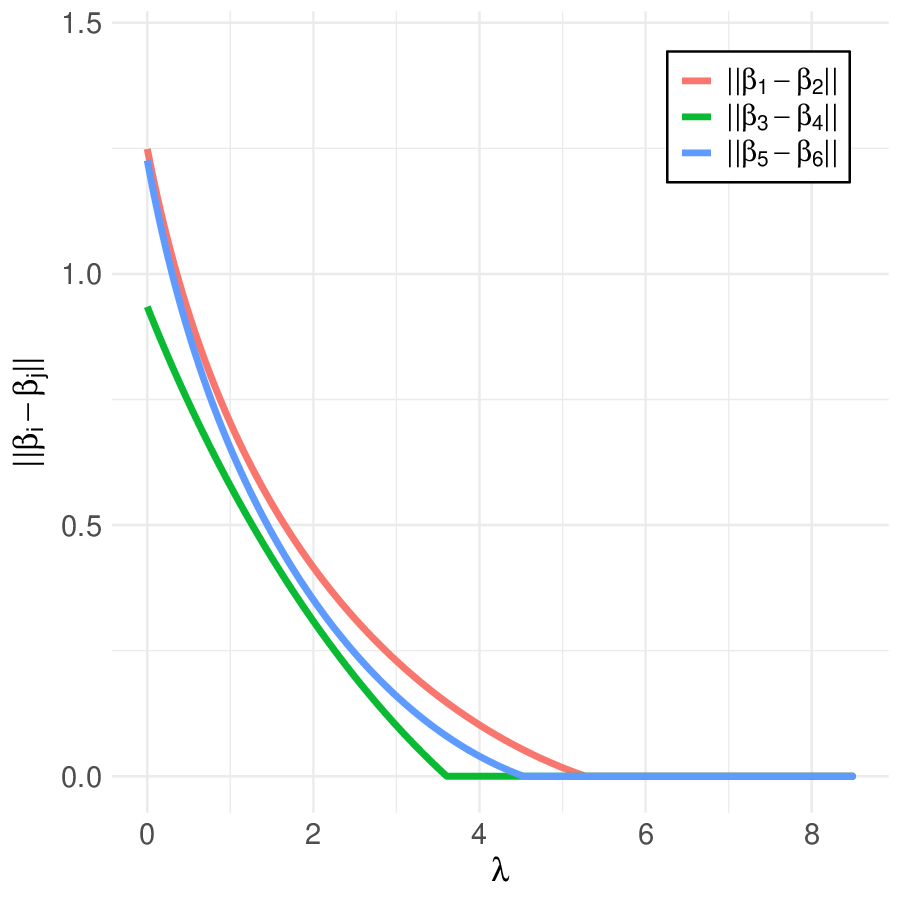}
    \end{minipage}
    \caption{Solution paths for the estimated pairwise differences of the coefficients in the pairwise vector fused LASSO and adaptive methods for Scenarios 1 and 2. The top panels correspond to Scenario 1, and the bottom panels represent Scenario 2. Left Panels show the results for the pairwise vector fused LASSO, and the Right panels show the results for the adaptive methods.}
    \label{fig:combined}
\end{figure}

Figure \ref{fig:combined} shows that while both PVF-LASSO and adaptive PVF-LASSO achieve the true grouping as $\lambda$ increases,  the latter achieves it much faster.

Tables \ref{tab:scenario1} and \ref{tab:scenario2} present the MSE for each of the five methods in Scenarios 1 and 2. In Scenario 1, the adaptive PVF-LASSO achieves the lowest MSE among all methods, and PVF-LASSO performs comparably with the Oracle and outperforms the Single and Separate methods. In Scenario 2, both PVF-LASSO and adaptive PVF-LASSO perform comparably with the Oracle and Separate. Tables \ref{tab:scenario1-betas} and \ref{tab:scenario2-betas} summarize the estimated $\beta$. The coefficients estimated by adaptive PVF-LASSO are closer to true values than those from the pairwise vector fused LASSO.

\begin{table}[ht]
\centering
\caption{Average test MSE (standard deviation) for Scenario 1 across 100 simulation replicates for each method.}
\small
\begin{tabular}{lcc}
\toprule
Method & $\sigma^2 = 1$ & $\sigma^2 = 5$ \\
\midrule
Oracle                        & 1.058 (0.188) & 5.293 (0.944) \\
Single                        & 3.974 (0.634) & 8.028 (1.268) \\
Separate                      & 1.099 (0.212) & 5.498 (1.063) \\
PVF-LASSO                     & 1.093 (0.211) & 5.464 (1.056) \\
Adaptive PVF-LASSO            & 1.050 (0.197) & 5.254 (0.985) \\
\bottomrule
\end{tabular}
\label{tab:scenario1}
\end{table}

\begin{table}[ht]
\centering
\caption{Average test MSE (standard deviation) for Scenario 2 across 100 simulation replicates for each method.}
\small
\begin{tabular}{lcc}
\toprule
Method & $\sigma^2 = 1$ & $\sigma^2 = 1/10$ \\
\midrule
Oracle                        & 0.352 (0.040) & 0.474 (0.047) \\
Single                        & 0.491 (0.046) & 0.499 (0.044) \\
Separate                      & 0.356 (0.046) & 0.475 (0.053) \\
PVF-LASSO                     & 0.363 (0.045) & 0.482 (0.048) \\
Adaptive PVF-LASSO            & 0.357 (0.048) & 0.478 (0.051) \\
\bottomrule
\end{tabular}
\label{tab:scenario2}
\end{table}

\begin{table}[ht]
\centering
\small
\caption{
Average estimated coefficients with standard errors (in parentheses) across 100 simulation replications for Scenario 1.
}
\begin{tabular}{llcccc} % Removed vertical | separators
\toprule
& \multirow{2}{*}{True $\beta$}
& \multicolumn{2}{c}{$\sigma^2=1$} % Removed |
& \multicolumn{2}{c}{$\sigma^2=5$} \\
\cmidrule(lr){3-4} \cmidrule(lr){5-6} % Keep cmidrule for header separation
& & PVF LASSO & Adaptive & PVF LASSO & Adaptive \\
\midrule 
& \phantom{-}1.0 & \phantom{-}0.963 (0.155) & \phantom{-}0.995 (0.110) & \phantom{-}0.963 (0.154) & \phantom{-}0.995 (0.110) \\
$\beta_1$ & \phantom{-}1.3 & \phantom{-}1.216 (0.144) & \phantom{-}1.273 (0.106) & \phantom{-}1.215 (0.144) & \phantom{-}1.273 (0.106) \\
& \phantom{-}1.3 & -1.226 (0.149) & -1.289 (0.113) & -1.226 (0.149) & -1.289 (0.113) \\
\addlinespace
& \phantom{-}1.0 & \phantom{-}0.956 (0.149) & \phantom{-}0.995 (0.111) & \phantom{-}0.956 (0.149) & \phantom{-}0.995 (0.111) \\
$\beta_2$ & \phantom{-}1.3 & \phantom{-}1.233 (0.138) & \phantom{-}1.273 (0.106) & \phantom{-}1.233 (0.138) & \phantom{-}1.273 (0.106) \\
& \phantom{-}1.3 & -1.247 (0.150) & -1.289 (0.111) & -1.247 (0.150) & -1.290 (0.111) \\
\addlinespace
& \phantom{-}0.5 & \phantom{-}0.461 (0.158) & \phantom{-}0.471 (0.123) & \phantom{-}0.461 (0.158) & \phantom{-}0.471 (0.123) \\
$\beta_3$ & -0.5 & -0.486 (0.151) & -0.510 (0.121) & -0.486 (0.151) & -0.511 (0.121) \\
& \phantom{-}0.5 & \phantom{-}0.467 (0.145) & \phantom{-}0.491 (0.112) & \phantom{-}0.467 (0.145) & \phantom{-}0.491 (0.112) \\
\addlinespace
& \phantom{-}0.5 & \phantom{-}0.424 (0.163) & \phantom{-}0.471 (0.123) & \phantom{-}0.424 (0.163) & \phantom{-}0.471 (0.123) \\
$\beta_4$ & -0.5 & -0.498 (0.165) & -0.510 (0.121) & -0.498 (0.165) & -0.511 (0.121) \\
& \phantom{-}0.5 & \phantom{-}0.475 (0.151) & \phantom{-}0.491 (0.112) & \phantom{-}0.475 (0.151) & \phantom{-}0.491 (0.112) \\
\addlinespace
& -1.5 & -1.438 (0.174) & -1.499 (0.115) & -1.438 (0.174) & -1.499 (0.115) \\
$\beta_5$ & -0.8 & -0.753 (0.142) & -0.787 (0.111) & -0.753 (0.142) & -0.787 (0.111) \\
& \phantom{-}0.8 & \phantom{-}0.746 (0.158) & \phantom{-}0.781 (0.116) & \phantom{-}0.746 (0.158) & \phantom{-}0.781 (0.116) \\
\addlinespace
& -1.5 & -1.417 (0.148) & -1.494 (0.117) & -1.417 (0.148) & -1.494 (0.117) \\
$\beta_6$ & -0.8 & -0.763 (0.151) & -0.792 (0.115) & -0.763 (0.151) & -0.792 (0.115) \\
& \phantom{-}0.8 & \phantom{-}0.753 (0.146) & \phantom{-}0.785 (0.122) & \phantom{-}0.753 (0.146) & \phantom{-}0.786 (0.122) \\
\bottomrule
\end{tabular}
\label{tab:scenario1-betas}
\end{table}

\begin{table}[ht]
\centering
\small
\caption{
Average estimated coefficients with standard errors (in parentheses) across 100 simulation replications for Scenario 2.
}
\begin{tabular}{llcccc}
\toprule
& \multirow{2}{*}{True $\beta$} 
& \multicolumn{2}{c}{$\sigma^2=1$} 
& \multicolumn{2}{c}{$\sigma^2=1/10$} \\
\cmidrule(lr){3-4} \cmidrule(lr){5-6}
& & PVF LASSO & Adaptive & PVF LASSO & Adaptive \\
\midrule
 & \phantom{-}1.0 & \phantom{-}0.966 (0.344) & \phantom{-}1.052 (0.381) & \phantom{-}0.744 (0.554) & \phantom{-}0.949 (0.680) \\
$\beta_1$ & \phantom{-}1.3 & \phantom{-}0.966 (0.344) & \phantom{-}1.052 (0.381) & \phantom{-}0.744 (0.648) & \phantom{-}1.245 (0.788) \\
 &    -1.3 & \phantom{-}1.242 (0.317) & \phantom{-}1.356 (0.391) & \phantom{-}0.988 (0.614) &    -1.155 (0.757) \\
\addlinespace
 & \phantom{-}1.0 & \phantom{-}0.929 (0.343) & \phantom{-}1.015 (0.386) & \phantom{-}0.669 (0.644) & \phantom{-}0.853 (0.788) \\
$\beta_2$ & \phantom{-}1.3 & \phantom{-}1.248 (0.347) & \phantom{-}1.363 (0.392) & \phantom{-}1.010 (0.603) & \phantom{-}1.276 (0.738) \\
 &    -1.3 &    -1.189 (0.347) &    -1.300 (0.392) &    -0.934 (0.625) &    -1.208 (0.748) \\
\addlinespace
 & \phantom{-}0.5 & \phantom{-}0.500 (0.276) & \phantom{-}0.525 (0.290) & \phantom{-}0.337 (0.608) & \phantom{-}0.459 (0.734) \\
$\beta_3$ &    -0.5 &    -0.465 (0.251) &    -0.485 (0.266) &    -0.286 (0.537) &    -0.360 (0.655) \\
 & \phantom{-}0.5 & \phantom{-}0.541 (0.252) & \phantom{-}0.564 (0.268) & \phantom{-}0.460 (0.522) & \phantom{-}0.574 (0.637) \\
\addlinespace
 & \phantom{-}0.5 & \phantom{-}0.468 (0.227) & \phantom{-}0.488 (0.237) & \phantom{-}0.309 (0.493) & \phantom{-}0.395 (0.590) \\
$\beta_4$ &    -0.5 &    -0.492 (0.263) &    -0.512 (0.276) &    -0.364 (0.661) &    -0.475 (0.790) \\
 & \phantom{-}0.5 & \phantom{-}0.472 (0.262) & \phantom{-}0.490 (0.276) & \phantom{-}0.339 (0.571) & \phantom{-}0.419 (0.687) \\
\addlinespace
 &    -1.5 &    -1.506 (0.351) &    -1.632 (0.406) &    -1.251 (0.595) &    -1.537 (0.705) \\
$\beta_5$ &    -0.8 &    -0.765 (0.289) &    -0.825 (0.319) &    -0.586 (0.569) &    -0.726 (0.682) \\
 & \phantom{-}0.8 & \phantom{-}0.759 (0.301) & \phantom{-}0.819 (0.335) & \phantom{-}0.673 (0.593) & \phantom{-}0.813 (0.723) \\
\addlinespace
 &    -1.5 &    -1.406 (0.310) &    -1.518 (0.356) &    -1.152 (0.564) &    -1.468 (0.674) \\
$\beta_6$ &    -0.8 &    -0.735 (0.310) &    -0.790 (0.338) &    -0.564 (0.609) &    -0.722 (0.732) \\
 & \phantom{-}0.8 & \phantom{-}0.733 (0.284) & \phantom{-}0.787 (0.313) & \phantom{-}0.616 (0.537) & \phantom{-}0.760 (0.660) \\
\bottomrule
\end{tabular}
\label{tab:scenario2-betas}
\end{table}

\section{Spotify music data}\label{sec:spotify}

We now apply our PVF-LASSO and adaptive PVF-LASSO to a data set obtained from the Spotify Web API, one of the most popular digital music streaming services. From Spotify for Developers (\url{https://developer.spotify.com/}), we extract information on songs, including their audio features (e.g., energy and danceability) and descriptive attributes (e.g., popularity, song name, genre, and subgenre) \citep{sciandra2022model}. Numerous studies have used the  Spotify data to predict music popularity based on various features; see, for example, \citet{sciandra2022model, terroso2023evolution, sandag2020predictive, gulmatico2022spotipred}.

In our application, we focus on popular songs with a popularity score ($Y$) greater than 68, resulting in a dataset of 487 songs. These songs are categorized into nine subgenres (represented by a categorical variable $U$): global ($U=1$), mainstream ($U=2$), soft ($U=3$), throwback ($U=4$), alternative ($U=5$), classic ($U=6$), spanish ($U=7$), 80s ($U=8$), and feel-good ($U=9$). We can further group these subgenres into broader genre categories: global, mainstream, soft, and throwback fall under pop, while alternative, classic, Spanish, 80s, and feel-good are categorized as rock. Our analysis considers six audio features as predictors: energy ($X_1$), danceability ($X_2$), loudness ($X_3$), liveliness ($X_4$), speechiness ($X_5$), and acousticness ($X_6$). The raw dataset is available at \url{https://www.kaggle.com/datasets/solomonameh/spotify-music-dataset/data}. Per subgenre category, we use 80\% of songs for training, a total of 386 songs, and the remainder to compute an estimated MSE on a test set of 101 songs. 

With these mixed predictors of subgenres and audio features, our goal is to model and predict song popularity while trying to collapse potentially similar subgenres. We compare the estimated coefficients and MSE of our two methods with three other methods: the Separate, the Single, and the Genre. As in the last section,     the Separate estimates a unique coefficient vector for each subgenre, resulting in nine coefficient vectors; the Single estimates a single coefficient vector, collapsing all subgenres into one group. The Genre, which plays the role of the Oracle in the simulation studies,  estimates two coefficient vectors by collapsing subgenres into broader genres—pop and rock. The Single and Genre methods serve as benchmarks for comparison depending on the context. The Single method collapses all categories, treating all subgenres simply as music, while the Genre method groups subgenres based on prior (albeit subjective) genre information, collapsing them accordingly. Unlike the Single and Genre methods, which impose fixed groupings, the PVF-LASSO and adaptive PVF-LASSO methods adaptively collapse subgenres without prior knowledge. For the PVF-LASSO and adaptive PVF-LASSO, we select the optimal $\lambda$ from a grid of values ranging from 0.01 to 0.5 in increments of 0.01, and we set $\gamma = 0.5$ for the adaptive PVF-LASSO. Because the sub-genres have significantly different sample sizes, we take the weighted version (\ref{eq:obj weighted}) of the objective function with $w \lo u = n \lo n \inv$. This prevents a large genre from dominating the optimization and thereby the category-collapsing process.

\begin{figure}
    \centering
    \includegraphics[width=0.7\linewidth]{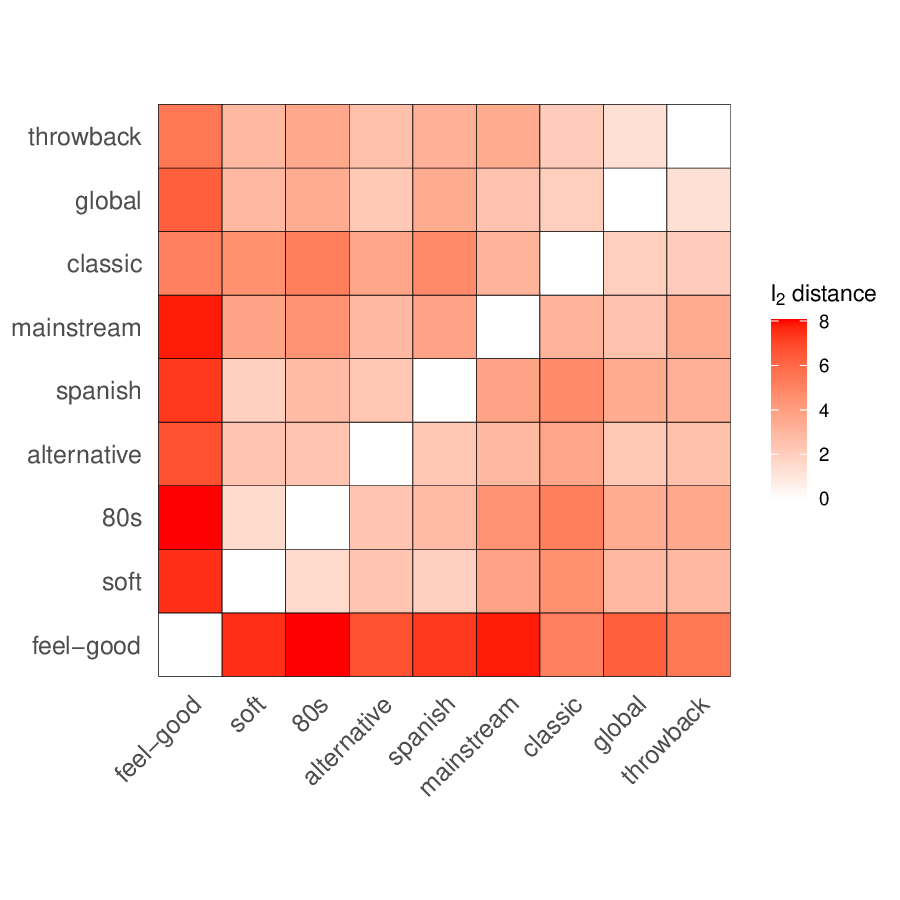}
    \caption{The pairwise $\ell_2$-distances between the coefficient vectors of each subgenre}
    \label{fig:l2idistance}
\end{figure}

Figure \ref{fig:l2idistance} displays the pairwise $\ell_2$-distances between the coefficient vectors of each subgenre of the Separate method. Notably, the feel-good subgenre stands out as significantly different from the others, suggesting that blindly collapsing all subgenres, as is done by the Single,  or collapsing them subjectively by pop versus rock, as is done by the Genre, may be inappropriate. 

Table \ref{tab:spotify} summarizes the estimated coefficient vectors and MSE for each method. The Separate method yields the highest MSE, indicating that performing regression individually for each subgenre is ineffective. The MSE of the Single method is lower than that of the Genre method, suggesting that collapsing subgenres based on genre can be misleading. This is because the feel-good subgenre differs from other subgenres within the rock genre than the pop genre. In this scenario, when comparing the Single and Genre methods, treating all subgenres as a single group is a better approach than relying on predefined genre-based groupings. The PVF-LASSO method merges seven subgenres—global, mainstream, soft, throwback, alternative, classic, and Spanish—resulting in three final groups: the merged group, 80s, and feel-good. The adaptive PVF-LASSO collapses eight subgenres—global, mainstream, soft, throwback, alternative, classic, Spanish, and 80s—forming just two groups: the merged group and feel-good. This collapsing is driven by the empirical regression structure. The grouped subgenres share similar regression coefficients across audio features. These similarities indicate that, in terms of popularity prediction, the subgenres are not statistically different.

Our methods stand in contrast to how genre and subgenre labels are typically assigned in practice. These labels are often predetermined by artists, producers, or marketing conventions, reflecting cultural identity or branding strategy rather than the actual acoustic or statistical profile of the music. Consequently, two subgenres might differ in name and intended audience but play nearly identical roles in a regression model. Our methods collapse subgenres with statistical evidence and reveal structural redundancy in the original labeling.

As shown in Table \ref{tab:spotify}, PVF-LASSO achieves the best MSE performance among all methods, followed by adaptive PVF-LASSO, Single, Genre, and Separate. These results show the advantage of adaptively collapsing subgenres, as it effectively reduces categorical complexity while improving the prediction of music popularity.

\begin{table}[htbp]
\centering
\caption{Estimated coefficient vectors across different methods. Vertical lines denote groups of coefficient vectors. $\beta_1$ for global, $\beta_2$ for mainstream, $\beta_3$ for soft, $\beta_4$ for throwback, $\beta_5$ for alternative, $\beta_6$ for classic, $\beta_7$ for spanish, $\beta_8$ for 80s, $\beta_9$ for feel-good.}
\footnotesize
\setlength{\tabcolsep}{4pt}
\begin{tabular}{l|c|c|c|c|c|c|c|c|c|c}
\hline
& $\beta_1$ & $\beta_2$ & $\beta_3$ & $\beta_4$ & $\beta_5$ & $\beta_6$ & $\beta_7$ & $\beta_8$ & $\beta_9$ & MSE \\
\hline
Separate & $-0.959$ & $-0.708$ & $\phantom{-}0.856$ & $-0.701$ & $\phantom{-}1.604$ & $\phantom{-}1.202$ & $-2.056$ & $-1.080$ & $\phantom{-}1.750$ & $32.004$ \\[-1pt]
         & $-0.204$ & $\phantom{-}0.652$ & $\phantom{-}0.450$ & $-0.282$ & $-0.130$ & $\phantom{-}0.427$ & $\phantom{-}0.483$ & $\phantom{-}2.020$ & $\phantom{-}0.859$ & \\[-1pt]
         & $\phantom{-}0.083$ & $-0.502$ & $-1.215$ & $\phantom{-}0.825$ & $-1.483$ & $-0.123$ & $\phantom{-}1.647$ & $\phantom{-}5.253$ & $-0.579$ & \\[-1pt]
         & $-0.175$ & $-1.096$ & $\phantom{-}0.204$ & $\phantom{-}0.678$ & $-0.491$ & $-0.385$ & $-0.239$ & $\phantom{-}1.777$ & $\phantom{-}1.116$ & \\[-1pt]
         & $-0.112$ & $\phantom{-}1.764$ & $\phantom{-}0.169$ & $-0.233$ & $-0.752$ & $-0.216$ & $\phantom{-}0.233$ & $-2.219$ & $\phantom{-}1.030$ & \\[-1pt]
         & $-0.171$ & $-1.175$ & $\phantom{-}1.591$ & $\phantom{-}0.416$ & $\phantom{-}1.379$ & $-0.460$ & $-0.369$ & $\phantom{-}0.477$ & $\phantom{-}0.575$ & \\
\hline
Single & \multicolumn{9}{c|}{$-0.264$} & $26.955$ \\[-1pt]
       & \multicolumn{9}{c|}{$\phantom{-}0.235$} & \\[-1pt]
       & \multicolumn{9}{c|}{$\phantom{-}0.273$} & \\[-1pt]
       & \multicolumn{9}{c|}{$\phantom{-}0.046$} & \\[-1pt]
       & \multicolumn{9}{c|}{$\phantom{-}0.127$} & \\[-1pt]
       & \multicolumn{9}{c|}{$\phantom{-}0.276$} & \\
\hline
Genre & \multicolumn{4}{c|}{$-0.427$} & \multicolumn{5}{c|}{$\phantom{-}0.058$} & $27.539$ \\[-1pt]
      & \multicolumn{4}{c|}{$\phantom{-}0.069$} & \multicolumn{5}{c|}{$\phantom{-}0.631$} & \\[-1pt]
      & \multicolumn{4}{c|}{$\phantom{-}0.093$} & \multicolumn{5}{c|}{$\phantom{-}0.474$} & \\[-1pt]
      & \multicolumn{4}{c|}{$\phantom{-}0.200$} & \multicolumn{5}{c|}{$-0.082$} & \\[-1pt]
      & \multicolumn{4}{c|}{$\phantom{-}0.120$} & \multicolumn{5}{c|}{$\phantom{-}0.032$} & \\[-1pt]
      & \multicolumn{4}{c|}{$\phantom{-}0.431$} & \multicolumn{5}{c|}{$\phantom{-}0.001$} & \\
\hline
PVF-LASSO & \multicolumn{7}{c|}{$-0.390$} & $-0.388$ & $-0.303$ & $26.544$ \\[-1pt]
         & \multicolumn{7}{c|}{$-0.018$} & $-0.021$ & $-0.032$ & \\[-1pt]
         & \multicolumn{7}{c|}{$\phantom{-}0.407$} & $\phantom{-}0.407$ & $\phantom{-}0.526$ & \\[-1pt]
         & \multicolumn{7}{c|}{$\phantom{-}0.022$} & $\phantom{-}0.027$ & $\phantom{-}0.022$ & \\[-1pt]
         & \multicolumn{7}{c|}{$\phantom{-}0.079$} & $\phantom{-}0.077$ & $\phantom{-}0.104$ & \\[-1pt]
         & \multicolumn{7}{c|}{$\phantom{-}0.136$} & $\phantom{-}0.136$ & $\phantom{-}0.083$ & \\
\hline
Adaptive & \multicolumn{8}{c|}{$-0.391$} & $-0.233$ & $26.547$ \\[-1pt]
         & \multicolumn{8}{c|}{$-0.018$} & $-0.046$ & \\[-1pt]
         & \multicolumn{8}{c|}{$\phantom{-}0.376$} & $\phantom{-}0.596$ & \\[-1pt]
         & \multicolumn{8}{c|}{$\phantom{-}0.024$} & $\phantom{-}0.021$ & \\[-1pt]
         & \multicolumn{8}{c|}{$\phantom{-}0.069$} & $\phantom{-}0.112$ & \\[-1pt]
         & \multicolumn{8}{c|}{$\phantom{-}0.134$} & $\phantom{-}0.036$ & \\
\hline
\end{tabular}
\label{tab:spotify}
\end{table}

\bibliographystyle{apalike}
\bibliography{biblio}

\end{document}